\documentclass{aa}  

\usepackage{graphicx}
\usepackage{subcaption}
\usepackage{txfonts}
\usepackage{color}
\usepackage{hyperref}
\usepackage{footnote}
\usepackage{longtable}
\usepackage{siunitx}
\makesavenoteenv{figure}
\usepackage{xcolor}
\usepackage{newtxtext,newtxmath}
\usepackage{amsmath}	

\DeclareRobustCommand{\VAN}[3]{#2}
\let\VANthebibliography\thebibliography
\def\thebibliography{\DeclareRobustCommand{\VAN}[3]{##3}\VANthebibliography}

\usepackage[normalem]{ulem}

\begin{document}

   \title{Stellar flares cannot explain the Galactic 511\,keV emission}


   \author{Saurabh Mittal\inst{1}\thanks{email: saurabh.mittal@uni-wuerzburg.de}
          \and
          Thomas Siegert\inst{1}
          \and
          Francesca Calore\inst{2}
          \and
          Fiona\,H. Panther\inst{3,4}
          \and
          Pasquale\,D. Serpico\inst{2}
          \and
          Hiroki Yoneda\inst{5,6,7}
          \and
          Niklas\,C. Bauer\inst{1}
          \and
          Tristan Bouchet\inst{1}
          \and
          Laura Eisenberger\inst{1}
          \and
          Mika\,A. Gelowicz\inst{1,8}
          \and 
          Tomohiko Oka\inst{1}
          \and
          Rudi Reinhardt\inst{1}
          \and
          Manuel\,R.\,H.\,W. Skalka\inst{1}
          \and
          Dimitris Tsatsis\inst{1}
          \and
          Manja\,L. Zimmerer \inst{1}
          }

   \institute{Julius-Maximilians-Universität Würzburg, Fakultät für Physik und Astronomie, Institut für Theoretische Physik und Astrophysik, Lehrstuhl für Astronomie, Emil-Fischer-Str.~31, D-97074 Würzburg, Germany
   \label{inst:jmu}
   \email{saurabh.mittal@uni-wuerzburg.de}
   \and
   Laboratoire d’Annecy de Physique Théorique (LAPTh), CNRS, USMB, 74940 Annecy, France
   \label{inst:lapth}
   \and
   Department of Physics, University of Western Australia, Crawley, WA 6009, Australia
   \label{inst:UWA}
   \and
   OzGrav: The ARC Centre of Excellence for Gravitational-wave Discovery, Australia
   \label{inst:GWD}
   \and
   Department of Physics, Kyoto University, Kitashirakawa Oiwake-cho, Sakyo-ku, Kyoto 606-8502, Japan
   \label{inst:kyoto}
   \and
   RIKEN Nishina Center, 2-1 Hirosawa, Wako, Saitama 351-0198, Japan
   \label{inst:riken}
   \and
   Kavli Institute for the Physics and Mathematics of the Universe (WPI), UTIAS, The University of Tokyo,
5-1-5 Kashiwanoha, Kashiwa, Chiba 277-8583, Japan
\label{inst:kavli}
\and
   Rheinische Friedrich-Wilhelms-Universität Bonn, Argelander-Institut für Astronomie (AIfA), Auf dem Hügel 71, D-53121 Bonn,
Germany
   \label{inst:bonn}
   }


\abstract
{The origin of the 511\,keV line signal in the Milky Way remains unresolved despite decades of observations. The measured flux of $\sim 3 \times 10^{-3}\,\mathrm{ph\,cm^{-2}\,s^{-1}}$ suggests a steady-state positron injection rate of $\sim 10^{43}$--$10^{44}\,\mathrm{e^+\,s^{-1}}$. One proposed contributor to this signal is stellar flaring activity since high energy Solar flares are known to produce positrons and associated annihilation radiation.}
{We estimate the quasi-persistent 511\,keV luminosity expected from flaring stellar populations and
estimate the Galactic contribution. We constrain the 511\,keV fluxes for different sources in the Galaxy, and in particular globular clusters.}
{Using Solar flare observations as a calibration baseline, we construct a hierarchical Bayesian model to link flare energy to 511\,keV luminosity. 
We further use flare frequency-energy distributions to estimate the time-averaged positron output of stellar populations. The resulting predictions are compared with INTEGRAL/SPI observations using spatial and population-based models.}
{We find that stellar flares fall short by several orders of magnitude in explaining the Galactic positron annihilation rate. 
Reproducing $\sim10\%$ of the observed luminosity in the Galactic bulge would require unphysically large maximum flare energies per star reaching up to $E_\mathrm{{max}} \gtrsim 10^{37-39}\,\mathrm{erg}$. Spatial modeling further shows that stellar flare scenarios cannot reproduce the observed 511\,keV morphology.}
{Stellar flares cannot be the dominant source of Galactic positrons. Although previous studies have shown that the measured 511\,keV morphology is broadly consistent with old stellar populations, our results exclude normal stellar flaring activity as the underlying source for this emission.} 
 
    %
    %
    %
    %


   \maketitle
%

\section{Introduction}\label{sec:intro}
The origin of the 511\,keV $\gamma$-ray line emission in the Milky Way has been a persistent enigma in astrophysics for more than five decades. 
This emission resulting from electron-positron annihilation to the 511\,keV line is reflected in a flux of $\sim 3 \times 10^{-3}\,\mathrm{ph\,cm^{-2}\,s^{-1}}$, corresponding to a positron injection rate of $\sim 10^{43}$--$10^{44}\,\mathrm{e^+\,s^{-1}}$
%
\citep[e.g.,][]{Skinner2014_511,Siegert2016_511, yoneda2025imaging, siegert2026possibility}.
Despite extensive research and observational efforts \citep[e.g.,][]{Jean2006_511, Churazov2005_511, Knoedlseder2005_511,Siegert2019_511lv, siegert2022measuring, das2025relaxation, knoedlseder2025}, the sources of these positrons remain unclear. 
The spatial morphology of this emission is one of the strongest constraints on the source population.
INTEGRAL/SPI \citep{Winkler2003_INTEGRAL, Vedrenne2003_SPI} observations show that the signal contains a bright central component associated with the Galactic bulge and nuclear region, together with a more extended disk \citep{Knoedlseder2005_511, Weidenspointner2008_511, Bouchet2010_511,Siegert2016_511}. 
This distribution is difficult to explain with sources that trace only recent star formation \citep{Prantzos11,Skinner2014_511, Siegert2023_511}, such as massive stars or core-collapse supernovae, because these are expected to be concentrated mainly in the Galactic disk.
Instead, the large bulge contribution motivates source populations associated with either old stars~\citep{Knoedlseder2005_511, Weidenspointner2008_511}, such as compact remnants or low-mass stellar populations, or scenarios in which positrons are produced elsewhere and transported before annihilation. 

%
%
A wide range of astrophysical sources have been proposed, including radioactive decay from nucleosynthesis products (e.g. $^{26}\mathrm{Al}$, $^{44}\mathrm{Ti}$, $^{56}\mathrm{Ni}$), compact objects such as pulsars and X-ray binaries, and more exotic scenarios such as dark matter~\citep[see][for a review]{Siegert2023_511}. 
%
%
%
Stellar flares provide another possible old-population channel. 
They are particularly interesting because low-mass stars are extremely numerous in the Milky Way and remain present in old stellar environments such as the Galactic bulge and globular clusters (GCs). 
Although individual stellar flares are transient on a timescale of minutes to hours, a large unresolved population of frequently flaring stars could, in principle, produce a quasi-persistent annihilation signal.
Solar flares have been known to produce positrons as a result of particle acceleration, $\beta^+$-unstable isotopes, pions, and subsequent decays. 
Annihilation of positrons in Solar flares mainly occurs through the formation of positronium (Ps), which decays either in three photons in the case of ortho-Ps, or into two photons (para-Ps) which results in the 511\,keV line.
%
%
Both components have been detected in high energy Solar flares by RHESSI, SMM, \textit{Fermi}-GBM, and other instruments \citep{Vestrand1999, Share2004_flares, Murphy2005_posiloss}.

\citet{Bisnovatyi-Kogan2017_511} suggested that stellar flares from the population of low-mass stars in the bulge could contribute significantly to the observed 511\,keV emission.
They estimate the positron production rate from one of the largest Solar flares and extrapolate this estimate to the flaring G-, K-, and M-type stars in the bulge.
A key assumption in their estimate is that nearly all flare-produced positrons escape before annihilating locally, leading to a much larger positron yield than implied by the observed Solar 511\,keV photons alone.
Their model also adopts the maximum flare energy from energetic flares observed in G-type stars with the Kepler telescope, while using Solar flares to set the lower energy bound because of detection limits.
%
%
%

The connection between stellar populations and the 511\,keV emission was further strengthened by \citet{Bartels2018_binaries511}, who proposed GCs as potential sources based on a correlation of the INTEGRAL/SPI 511\,keV measurements with the \textit{Fermi}/LAT $\gamma$-ray excess at GeV energies. 
This connection aligns well with the stellar flare hypothesis, as GCs contain high concentrations of old, low-mass stars known to produce frequent flares.

While these studies provide compelling reasons to consider stellar flares as a source of Galactic positrons, a comprehensive analysis of this scenario is still missing. 
In this work, we test the hypothesis that stellar flares can explain the Galactic 511\,keV emission.
Our approach combines two complementary methods.
First, we construct a hierarchical model for the quasi-persistent 511\,keV luminosity of stellar populations, calibrated using Solar flare observations and flare frequency-energy distributions across stellar types.
Second, we test the resulting stellar flare morphology directly against 20 years of INTEGRAL/SPI observations. 
%

This paper is structured as follows:
In Sect.\,\ref{sec:hierarchical_model}, we introduce the stellar flare hypothesis and the hierarchical framework used in this work.
In Sect.\,\ref{sec:flaring_sun}, we describe Solar flare measurements that allow us to construct a correlation between the flare energy and the 511\,keV luminosity.
We extend this correlation to form a model for the quasi-persistent 511\,keV luminosity of different stellar types in Sect.\,\ref{sec:star_groups}.
In Sect.\,\ref{sec:luminosity_predictions}, we estimate the total Galactic 511\,keV luminosity expected from stellar flares, and calculate expected luminosities for all known GCs of the Milky Way.
%
In Sect.\,\ref{sec:INTEGRAL/SPI}, we fit the stellar flare model for different stellar types to the observed data from INTEGRAL/SPI.
%
%
We discuss our results in Sect.\,\ref{sec:discussion}, including systematic uncertainties and model assumptions.
Finally, we give a summary and an outlook in Sect.\,\ref{sec:conclusion}.
%


%
\section{Stellar flare hypothesis for Galactic positrons}\label{sec:hierarchical_model}
%
%
An intermittently flaring stellar population produces a quasi-persistent positron annihilation signal over time.
%
The total 511\,keV luminosity of a flaring star can be estimated by integrating the contribution of flares of different energies weighted by their occurrence frequency and duration.
%
%
The quasi-persistent 511\,keV luminosity of a star of mass $M_\mathrm{*}$ is expressed as:
\begin{equation}
    \begin{aligned}
        \mathrm{L^{*}_{511,QP}} (\mathrm{M_*}) = \int_{\mathrm{E_{min}}}^{\mathrm{E_{max}}}\mathrm{L_{511}^\odot}(\mathrm{E_F})\,\nu(\mathrm{E_F, M_*})\,\Delta\mathrm{t}(\mathrm{E_F, M_*})\,d\mathrm{E_F}\mathrm{,}
    \end{aligned}
    \label{eq:quasi_persistent_all_stars}
\end{equation}
where $\mathrm{L_{511}^\odot}(\mathrm{E_F})$ is the 511\,keV luminosity of a flare of energy $\mathrm{E_F}$ calibrated from Solar observations~(Sect.\ref{sec:flaring_sun}), $\nu(\mathrm{E_F, M_*})$ is the flare frequency-energy distribution (FFD) for a star of mass $M_*$ (Sect.\,\ref{sec:star_groups}), and $\Delta\mathrm{t}(\mathrm{E_F, M_*})$ is the flare duration (Sect.\,\ref{subsec:flare_duration}).
%
%
The total Galactic luminosity is then obtained by integrating over the Galactic stellar population,
\begin{equation}
    \mathrm{L}_{511}^{\mathrm{MW}} = \int  \mathrm{L^{*}_{511,QP}} (\mathrm{M_*})~\Phi(\mathrm{M_*})~\,d\mathrm{M_*},
    \label{eq:luminosity_MW}
\end{equation}
where $\Phi(\mathrm{M_*})$ describes the Galactic present-day mass function \citep[PDMF;][]{chabrier2003galactic}.
The observationally constrained parameter distributions are explained in the following.
%
%

\section{The flaring Sun}\label{sec:flaring_sun}
\subsection{Solar flare observables and datasets}\label{subsec:sun_datasets}
To establish a relationship between flare energy and 511\,keV luminosity, we use Solar flares as the calibration baseline.
This choice is motivated by the fact that the Sun is the only star for which $\gamma$-ray line emission from individual flares has been measured directly.
We examine several intermediate correlations to construct this relation.
This step-by-step approach is necessary due to the scarcity of direct measurements linking flare energy to 511\,keV emission.
By linking and chaining these correlations, we can construct a robust model that predicts the 511\,keV luminosity as a function of flare energy.
The correlations are derived from different instruments (see below) and datasets, each with its own systematic uncertainties.
These are incorporated within our hierarchical framework.
\subsection{Modeling correlations with unknown uncertainties}\label{sec:stan}
In our analysis of Solar flare data, we encounter several challenges related to data quality and uncertainty quantification.
To address these issues, we utilise \texttt{stan}\footnote{https://mc-stan.org}, which is a probabilistic modeling language. 
The advantage of \texttt{stan} in this context is that it naturally incorporates the required tools for constructing hierarchical models with vague uncertainties.
%
%
Our primary model structure consists of power law relationships, as suggested by both the observed shape of correlations in our data and previous works in the field~\citep[e.g.,][]{Aschwanden2015, Shih2009}.
The general form of a power-law model is
\begin{equation}
\label{eq:power_law}
    y = \mathcal{N} \times x^\eta
\end{equation}
where $y$ is the dependent variable, $x$ is the independent variable, $\mathcal{N}$ is the normalization, and $\eta$ is the power-law index.
The main challenge we face with Solar flare literature data are missing uncertainties for many measurements.
%
%
To address this, we implement a hierarchical model structure that allows us to estimate these uncertainties as part of the fitting process. We assume that the true uncertainties follow a log-normal distribution
\begin{equation}
\label{eq:normal_dist}
    \mathrm{log}(\sigma_{\mathrm{i}}) \sim \mathrm{Normal}(\mu_{\sigma},s_{\sigma})\mathrm{,}
\end{equation}
where $\sigma_{\mathrm{i}}$ is the uncertainty in the $i$-th data point, $\mu_{\sigma}$ is the mean log-uncertainty, and $s_{\sigma}$ is the standard deviation of the log-uncertainties.
This approach allows us to capture both measurement uncertainties and intrinsic scatter in the relationships we are modeling.
Furthermore, we recognise that many of the correlations we observe may be influenced by hidden variables not explicitly included in our models or provided by measurements. 
These include factors such as magnetic field strength, the directional nature of the peak flux measured by the Geostationary Operational Environmental Satellite (GOES), or the phase of the Solar cycle.
To account for these potential sources of systematic uncertainties, we introduce a multiplicative term to each uncertainty that we also estimate as part of our model fitting process.
The latent values (red markers) shown in Figs.\,\ref{fig:aschw_energy_peak_flux_stanfit}, \ref{fig:shih_rhessi_stanfit}, and \ref{fig:smm_lum_511_direct_stanfit} represent the inferred true data points after this uncertainty treatment.
These are then used to constrain the power-law correlations in a single inference step.
%
%
We use weakly informative priors for all our model parameters to regularise our fits and prevent over-fitting. 
For example:
\begin{equation}
\begin{aligned}
\label{eq:priors}
    \mathcal{N} &\sim \mathrm{Normal}(\mu_{\mathcal{N}}, \sigma_{\mathcal{N}}) \\
     \eta &\sim \mathrm{Normal}(\mu_{\eta}, \sigma_{\eta}) \\
      \sigma_{\mathrm{sys}} &\sim \mathrm{Cauchy}(1.0, 1.0)
\end{aligned}
\end{equation}
The values for $\mu_{\mathcal{N}}$ and $\mu_{\eta}$ are estimated using a first order approximative power-law fit to the data.
The values for $\sigma_{\mathcal{N}}$ and $\sigma_{\eta}$ are allowed a wide range to not force fit the data.
For sampling, we use 
the No-U-Turns sampler (NUTS), 
typically running two chains for 1000 iterations each. 
%
We assess convergence using the $\hat{R}$ statistic, ensuring $\hat{R}$ values consistent with unity.
This approach allows us to robustly estimate correlations between Solar flare parameters even in the presence of unknown uncertainties and potential systematic effects, providing a solid foundation for subsequent analysis and model development.

%


\subsection{Correlation chain}\label{subsec:correlation_chain}
\subsubsection{Flare class conversion to peak soft X-ray flux}\label{subsec:class_to_flux}
The GOES flare classification system provides a standardized way to categorize Solar flares based on their peak soft X-ray flux in the 1--8\,\AA\, band.
The classification follows a logarithmic scale, with each class letter (A, B, C, M, X) representing an order of magnitude increase in peak flux.
Additionally, each class is further subdivided into nine linear levels which is denoted by a number from 1--9 following the class letter. 
%
Since most available datasets for Solar flare observations lists only the flare class, this standardised conversion factor was used to convert flare class to peak soft X-ray flux. 

\subsubsection{Flare energy vs. peak soft X-ray flux}
\begin{figure}[!t]
    \centering
    \includegraphics[width=0.95\columnwidth]{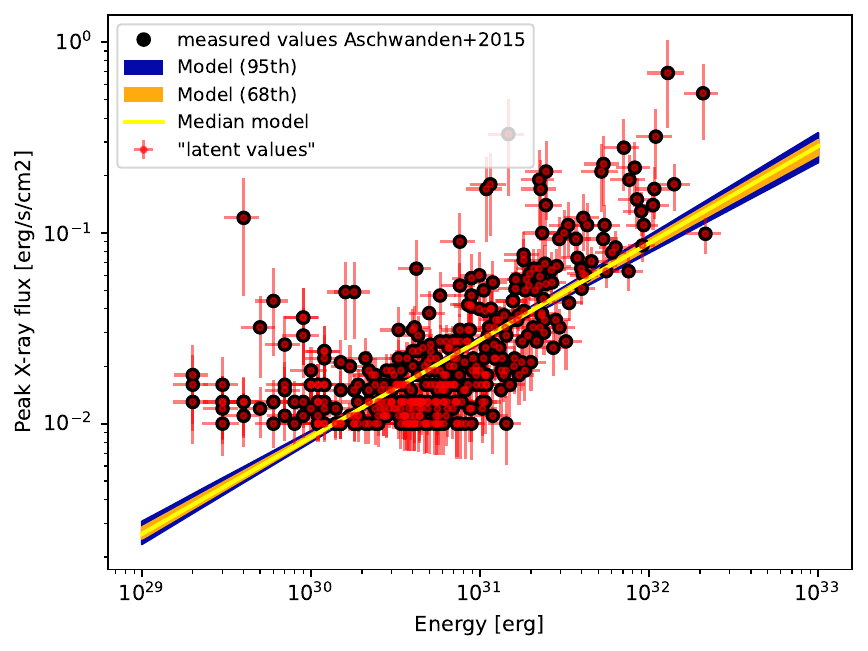}
    \caption{Power-law fit for flare energy vs. peak soft X-ray flux. Measured values (black) are obtained from \citet{Aschwanden2015}. 
    The red error bars 
    are the values that were used for fitting after assigning a systematic uncertainty to the data to find a robust correlation value since the measured data does not have uncertainties. The yellow line shows the median model 
    with the 68th and 95th percentiles shown as orange and blue band.}
    \label{fig:aschw_energy_peak_flux_stanfit}
\end{figure}

To correlate the total flare energy with the peak soft X-ray flux, we use data from \citet{Aschwanden2015}, that includes roughly 400 M- and X-class flares observed with the Atmospheric Imaging Assembly (AIA) onboard the Solar Dynamics Observatory (SDO). 
Our model is
\begin{equation}
\label{eq:SXR_energy}
    \mathrm{F_{SXR}} = \mathrm{A} \times \left(\frac{\mathrm{E_F}}{\mathrm{E_A}}\right)^\alpha\mathrm{,}
\end{equation}
where $\mathrm{F_{SXR}}$ is the peak soft X-ray flux in $\mathrm{erg\,cm^{-2}\,s^{-1}}$, $\mathrm{A}$ is the normalisation constant, $\mathrm{E_A}$ is the pivot energy, and $\alpha$ is the power-law index.
We obtain best-fit values of $\mathrm{A} = (8.5 \pm 0.2) \times 10^{-3}\,\mathrm{erg\,cm^{-2}\,s^{-1}}$, $\alpha = 0.51 \pm 0.02$, for pivot energy $\mathrm{E_A} = 10^{30}\,\mathrm{erg}$. 
Fig.\,\ref{fig:aschw_energy_peak_flux_stanfit} shows the best fit power-law relationship to this data, along with the assumed uncertainties.
The narrow width of the 95th percentile in Fig.\,\ref{fig:aschw_energy_peak_flux_stanfit} reflects that the average trend is relatively well constrained despite the large individual flare scatter.
This is due to the large sample size of this dataset that densely populates over a relatively narrow energy range.
The 95th percentile band does not provide a prediction interval for individual flare measurements but rather a posterior prediction after assuming uncertainties and intrinsic scatter.

\subsubsection{Peak soft X-ray flux vs. 2.223\,MeV line flux}
\begin{figure}[!t]
    \centering
    \includegraphics[width=0.95\columnwidth]{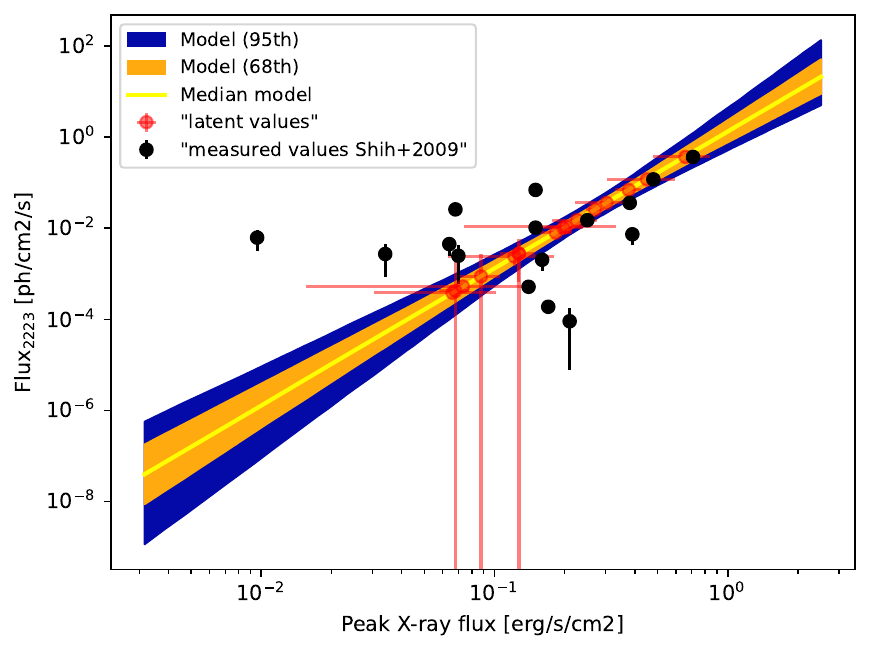}
    \caption{Same as Fig.\,\ref{fig:aschw_energy_peak_flux_stanfit} but for a power-law fit for peak soft X-ray flux vs. measured 2.223\,MeV line flux for different flares. Data points (black) have been obtained from \citet{Shih2009}.}
    %
    \label{fig:shih_rhessi_stanfit}
\end{figure}

The 2.223\,MeV $\gamma$-ray line is produced in Solar flares after neutrons become thermalized in the photosphere and are captured by protons resulting in the production of deuterium 
with a binding energy of 2.223\,MeV. 
\citet{Shih2009} shows a close correlation between the neutron capture line and the bremsstrahlung continuum above 300\,keV based on several flares observed with RHESSI. 
We use data from this study and correlate the 2.223\,MeV emission to the peak soft X-ray flux. 
Our model reads
\begin{equation}
\label{eq:2223_SXR}
    \mathrm{F_{2223}} = \mathrm{B} \times \mathrm{F_{SXR}}^\beta\mathrm{,}
\end{equation}
where $\mathrm{F_{2223}}$ is the 2.223\,MeV line flux in $\mathrm{ph\,cm^{-2}\,s^{-1}}$, $\mathrm{B}$ is the normalisation constant, $\mathrm{F_{SXR}}$ is the peak soft X-ray flux in $\mathrm{erg\,cm^{-2}\,s^{-1}}$, and $\beta$ is the power-law index.
We obtain best-fit values of $\mathrm{B} = (1.35 \pm 0.74)$, and $\beta = (3.02 \pm 0.34)$.
Fig. \ref{fig:shih_rhessi_stanfit} shows the best fit power-law relationship to this data, along with the measured (black) and fitted (red) uncertainties.
We can see that the fit is highly influenced by the four strongest flares that have the smallest uncertainties and 2.223\,MeV fluxes, $\geq 10^{-2}\,\mathrm{ph\,cm^{-2}\,s^{-1}}$.

\subsubsection{2.223\,MeV line flux vs. 511\,keV line flux}
\begin{figure}[!t]
    \centering
    \includegraphics[width=0.95\columnwidth]{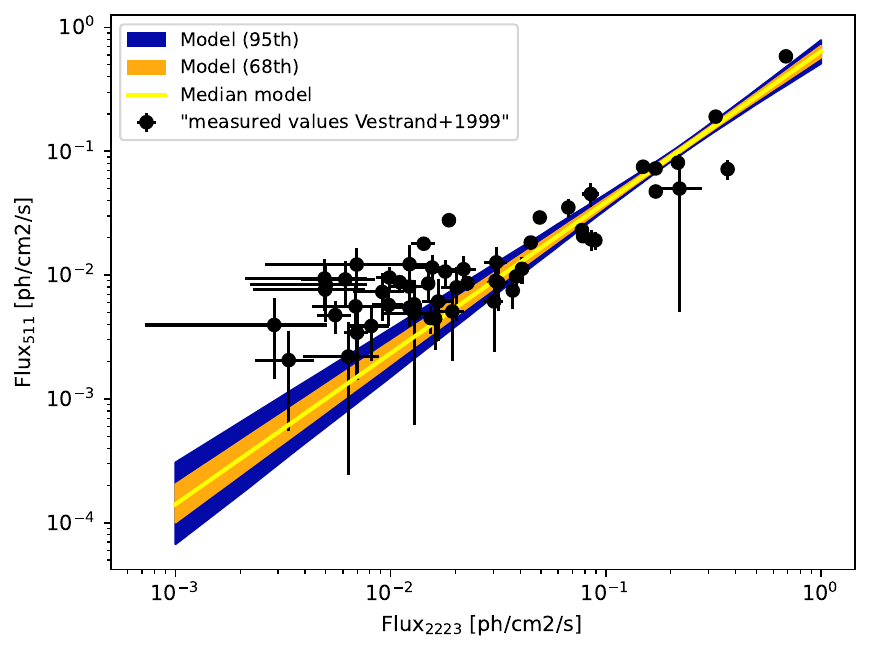}
    \caption{Same as Fig.\,\ref{fig:aschw_energy_peak_flux_stanfit} but for a power-law fit for 2.223\,MeV line flux vs. 511\,keV flux for different flares using data from \citet{Vestrand1999}.}
    %
    \label{fig:smm_2223_511_stanfit}
\end{figure}

Finally, we correlate the 2.223\,MeV line flux with the 511\,keV line flux.
For this, we use data from \citet{Vestrand1999} that contains $\sim200$ Solar flares detected above 300 keV by the Gamma Ray Spectrometer (GRS) onboard the Solar Maximum Mission (SMM) satellite.
This sample was obtained during 1980--1989 covering the latter half of the 21st Solar sunspot cycle.
The large duration of this dataset allows us to monitor flares during both the strong and the weak activity cycle of the Sun.
We again establish a power-law correlation,
\begin{equation}
\label{eq:511_2223}
    \mathrm{F_{511}} = \mathrm{C} \times \mathrm{F_{2223}}^\gamma\mathrm{,}
\end{equation}
where $\mathrm{F_{511}}$ is the 511\,keV line flux, $\mathrm{C}$ is the power-law normalisation, $\mathrm{F_{2223}}$ is the 2.223\,MeV line flux, and $\gamma$ is the power-law index.
We obtain best-fit values of $\mathrm{C} = (0.65 \pm 0.07)$, and $\gamma = (1.22 \pm 0.07)$, and the fit is shown in Fig.\,\ref{fig:smm_2223_511_stanfit}.
%
%
 
\subsection{Predicting 511\,keV luminosity from flare energy}\label{subsec: indirect_fit}
\begin{figure}[!t]
    \centering
    \includegraphics[width=0.95\columnwidth]{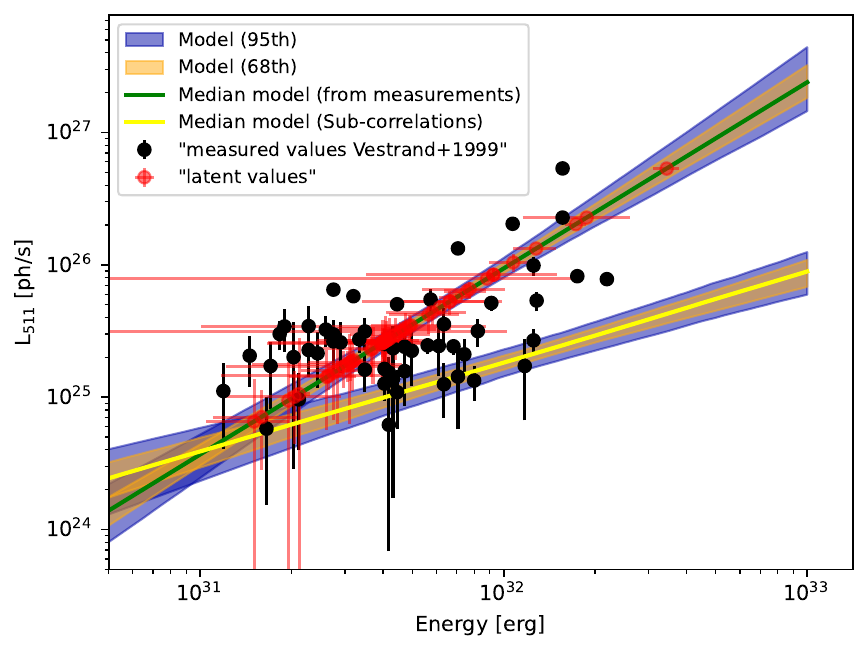}
    \caption{
    Final relation expected for the 511\,keV luminosity of the Sun as a function of flare energy (yellow line)
    It is obtained by jointly fitting the correlations shown in Figs.\,\ref{fig:aschw_energy_peak_flux_stanfit}, \ref{fig:shih_rhessi_stanfit}, and \ref{fig:smm_2223_511_stanfit}. Measured values from \citet{Vestrand1999} are only shown as a reference.
    To establish consistency in measured data vs. our estimated relation, 
    alternatively, the green line shows the median model obtained from fitting directly to the measured SMM data.
    Uncertainty bands are the same as in previous figures.} 
    %
    \label{fig:smm_lum_511_direct_stanfit}
\end{figure}

Combining the above sub-correlations (Eqs.\,\ref{eq:SXR_energy}, \ref{eq:2223_SXR}, and \ref{eq:511_2223}), we can now predict the 511\,keV line luminosity for a given flare energy for the Sun by
%
%
\begin{equation}
\label{eq:511_energy}
\begin{aligned}
    \mathrm{L_{511}^\odot}(\mathrm{E_F}) &= 4 \times \pi \times \mathrm{d}^2_\odot \times \mathrm{C} \times \mathrm{B}^\gamma \times \mathrm{A}^{\beta\gamma} \times \left(\dfrac{\mathrm{E_F}}{\mathrm{E_A}}\right)^{\alpha\beta\gamma} \\
    &\equiv \mathrm{D} \times \left(\dfrac{\mathrm{E_F}}{\mathrm{E_A}}\right)^\delta = 8.0 \times 10^{23} \left(\dfrac{\mathrm{E_F}}{10^{30}}\right)^{0.68}\,\mathrm{ph\,s^{-1}.}
\end{aligned}
\end{equation}
Fitting all the sub-correlation jointly results in best-fit values of $\mathrm{D} = (8.0 \pm 3.4) \times 10^{23}\,\mathrm{ph\,s^{-1}}$ and $\delta = 0.68 \pm 0.08$.
The joint fit simultaneously constraints the three empirical sub-correlations.
Each relation is modeled as a power-law with their individual fit parameters while the final relation for 511\,keV flux vs. flare energy is obtained by propagating through the chain of correlations.
This allows for the uncertainties in the intermediate correlations to be consistently carried into the final fit.
Fig.\,\ref{fig:smm_lum_511_direct_stanfit} shows the joint fit for the three sub-correlations.
For reference, the figure also shows the 511\,keV luminosity of Solar flares as directly measured \citep{Vestrand1999}.
%
%

\subsection{Estimating 511\,keV flux directly from measurements} \label{subsec: direct_fit}
\citet{Vestrand1999} derived spectral parameters, such as positron annihilation line fluence and annihilation continuum fluence for $\gamma$-ray flares observed by SMM, for flares of different GOES class. 
Flare class to X-ray peak flux conversion for these flares was performed as described in Sect.\,\ref{subsec:class_to_flux}.
\citet{kretzschmar2011sun} calculated the total Solar irradiance using white light flares which can be characterized by a scaling law relationship between the bolometric energy $\mathrm{E_{\mathrm{bol}}}$ in $\mathrm{erg}$ and the GOES 1--8\,\AA\, peak soft X-ray flux $\mathrm{F_{\mathrm{SXR}}}$ in units of $\mathrm{W\,m^{-2}}$ given by

\begin{equation}
    \label{eq:bol_to_flux}
    \dfrac{\mathrm{E_{\mathrm{bol}}}}{10^{30}\,\mathrm{erg}} \approx \left( \dfrac{\mathrm{F_{\mathrm{SXR}}}}{2 \times 10^{-6}\,\mathrm{W\,m^{-2}}}\right)^{0.78}\mathrm{.}
\end{equation}
We use Eq.\,\ref{eq:bol_to_flux} to convert the peak soft X-ray flux to the bolometric energy of the flare and correlate it with the 511\,keV flux.
Fig.\,\ref{fig:smm_lum_511_direct_stanfit} shows the best-fit power-law relation to this data (converted to luminosity).
Since, the bolometric energy did not have any measurement uncertainty, we estimate the uncertainty in the data in the fitting process.
We obtain best-fit values for $\delta'= (1.41 \pm 0.10)$ and $\mathrm{D'}= (1.4 \pm 0.6) \times 10^{23}\,\mathrm{ph\,s^{-1}}$ with a pivot energy of $10^{30}\,\mathrm{erg}$.
This is substantially steeper than our sub-correlation treatment so that both methods describe a systematic uncertainty range.

%
%

\section{Flare frequency distributions and stellar populations}\label{sec:star_groups}
To examine the total contribution of stellar flares to the Galactic 511\,keV emission, we must consider the diverse stellar population in our Galaxy. 
While our initial analysis uses Solar flares as a template, different stellar classes exhibit varying flaring behaviors. 
We begin by examining two distinct approaches:
1) Solar model: Assuming all stars behave like the Sun.
2) Stellar-class specific model: Accounting for varying flaring activity in different stellar classes, particularly M- and K-type stars.

\subsection{Flare frequency distribution for the Sun}
The FFD of Solar flares follows a power-law relationship across a broad range of energies, from so-called nano-flares ($10^{24}$--$10^{26}\,\mathrm{erg}$) to super-flares ($10^{34}$--$10^{36}\,\mathrm{erg}$) \citep{aschwanden2000, maehara2012superflares}:
\begin{equation}
\label{eq:power_law_flare_frequency}
    \nu(\mathrm{E_F, \odot}) = \dfrac{\mathrm{dN}}{\mathrm{dE\,dt}} = \mathrm{A_\nu}  \times \left(\dfrac{\mathrm{E_F}}{\mathrm{E_{A_\nu}}}\right)^{\alpha_\nu},
\end{equation}
where, 
\begin{itemize}
    \item[$\bullet$] Nanoflares: $\alpha_\nu \approx -1.79 \pm 0.08$ \citep{aschwanden2000} 
    \item[$\bullet$] Microflares: $\alpha_\nu \approx -1.74$ \citep{shimizu1995}
    \item[$\bullet$] Solar flares: $\alpha_\nu \approx -1.53 \pm 0.02$ \citep{crosby1993}
    \item[$\bullet$] Superflares: $\alpha_\nu \approx -2.3 \pm 0.3$ \citep{maehara2012superflares}
\end{itemize}
and $\mathrm{A_\nu}$ is the differential frequency of a flare of a given energy in units of $\mathrm{erg^{-1}\,s^{-1}}$.
This relationship for flares from Sun-type stars was built based on individual studies from \citet{aschwanden2000} for extreme-ultraviolet nano-flares using the \textit{Transition Region And Coronal Explorer (TRACE)}, \citet{shimizu1995} for micro-flares using the Soft X-ray telescope aboard the Yohkoh satellite, \citet{crosby1993} analyses of over 12000 Solar flares recorded with the Hard X-ray Burst Spectrometer on SMM, and super-flares on slowly rotating main sequence G-type stars similar to the Sun by \citet{maehara2012superflares}.
Fig.\,\ref{fig:flare_freq_energy_all_stellar_types} shows the FFD for the Sun and different types of flares recorded from the Sun or Sun-like stars.
The largest flares recorded from the Sun over the past few decades have reached energies of $\sim 10^{33}\,\mathrm{erg}$ \citep{schrijver2012estimating, aulanier2013}.
Based on historical reports for sunspot and Solar active region properties in the photosphere, \citet{aulanier2013} places an upper limit on Solar flare energy of only $\sim 6 \times 10^{33}\,\mathrm{erg}$.
We note that Sunspot groups larger than historically reported would yield stronger flares for the Sun, and \citet{yang2019flare} estimate that a super-flare with energy $\sim 10^{34}\,\mathrm{erg}$ occurs on the Sun once in 5500\,yr.
Furthermore, flares in active Sun-like stars reach energies up to $\sim 10^{36-37}\,\mathrm{erg}$ \citep{maehara2012superflares}.
These solar FFDs provide the baseline against which the more active stellar populations are compared. 
They also illustrate why the high-energy end of the flare distribution is critical for the 511\,keV problem: the positron yield per flare increases with energy, but the occurrence rate decreases steeply. 
The total quasi-persistent luminosity,
\begin{equation}
    \mathrm{L}_{511,\mathrm{QP}}^*(\mathrm{M}_*) = \frac{\mathrm{D\,A_\nu\,T}}{1 + \epsilon}\mathrm{E_A}^{-\epsilon} [{\mathrm{E_{max}}}^{1+\epsilon} - {\mathrm{E_{min}}}^{1+\epsilon} ]\mathrm{,}
    \label{eq:qp_lumi}
\end{equation}
with $\epsilon = \delta + \alpha_\nu + \tau$ (see Sects.\,\ref{sec:FFD_stars} \& \ref{subsec:flare_duration} for $A_\nu$ and $T$) is therefore sensitive to the FFD slope, the normalisation, and the assumed maximum flare energy. 
%
%
%
The luminosity is further controlled by the highest energy flares when $\epsilon > -1$, which is true in most populations, and by the lowest energy flares when $\epsilon < -1$ (possible for A-type stars).

\subsection{Flare frequency distribution for different stellar types}\label{sec:FFD_stars}
\begin{figure}[!t]
    \centering
    \includegraphics[width=\columnwidth]{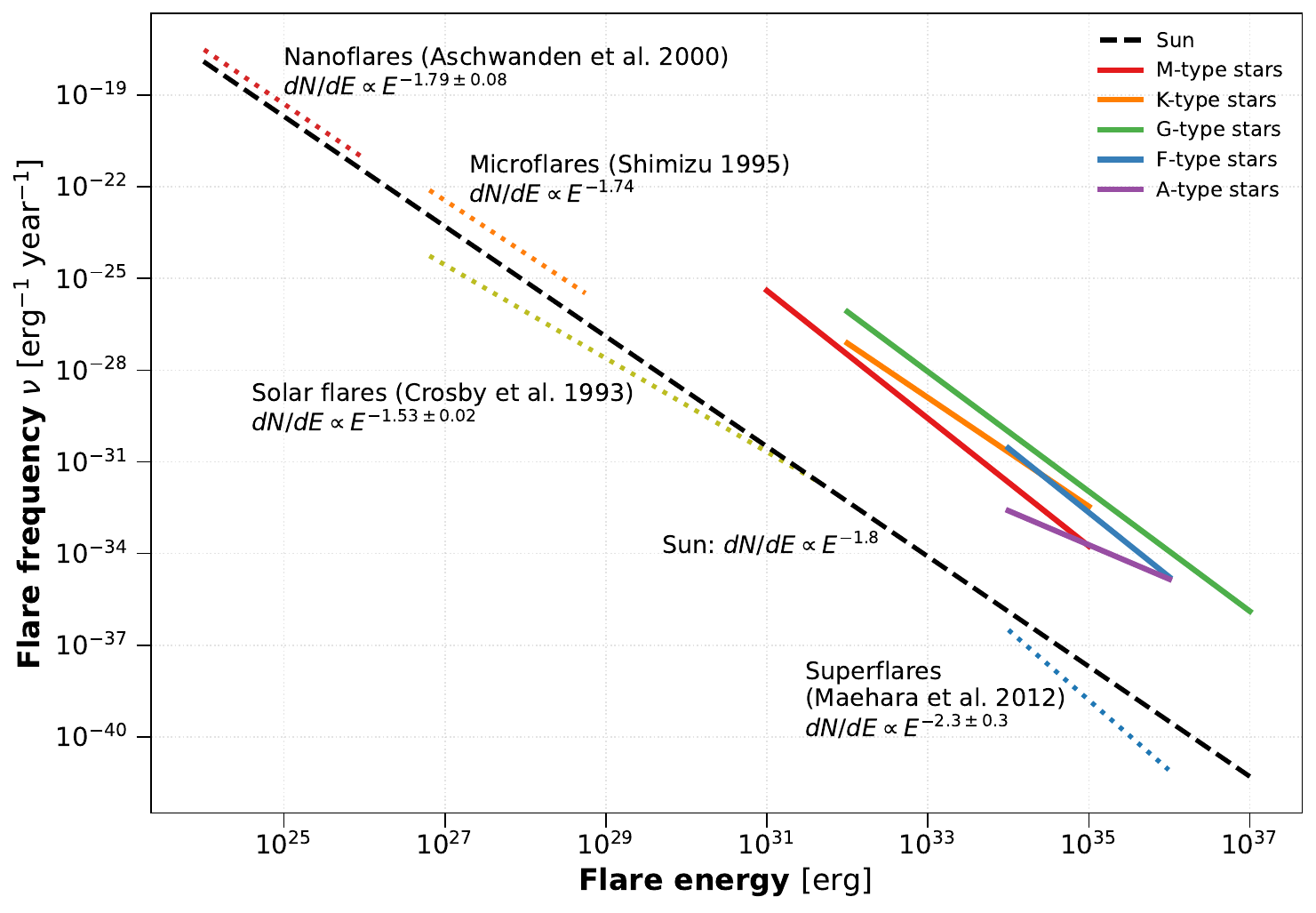}
    \caption{Flare frequency-energy relation for different stellar types~\citep{yang2019flare}. The Sun shows weaker activity compared to other stellar types and to other G-type stars. The low energy activity of other stellar types is also unknown due to detection limits. }
    \label{fig:flare_freq_energy_all_stellar_types}
\end{figure}

Stellar flare activity varies strongly with stellar type. 
Cool stars, especially K and M dwarfs, are generally more magnetically active than the present-day Sun and show higher flare occurrence rates with higher energy output~\citep{walkowicz2011white}. 
This enhanced activity is also closely connected to rotation. 
In the Kepler flare catalogue of~\citet{yang2019flare}, most flare stars are rapid rotators with approximately 70\% having rotation periods shorter than 10 days, and approximately 95\% having periods shorter than 30 days. 
The same study found that the FFDs of flaring stars from F to M spectral types are broadly described by power laws with indices \(\alpha_\nu\simeq-2\). 
Observations from Kepler, TESS, and NGTS generally support steep FFDs for active stars. 
\citet{jackman2021NGTS} measured FFDs for stars spanning approximately F8--M6 using NGTS and found power-law slopes broadly consistent with those inferred from Kepler and TESS studies. 
Studies of M dwarfs with TESS also find slopes in the approximate range \(\alpha_\nu\simeq-1.7\) to \(-2.2\) \citep{gunther2020TESS,gao2022correcting}. 
This suggests that the flare-generation process is broadly similar across these stellar classes, although the flare rates and maximum energies differ~\citep{yang2019flare}.\\
The Sun is therefore a relatively quiet reference case. 
Other G-type stars can flare more frequently than the present-day Sun~\citep{maehara2012superflares}, and active K- and M-type stars can produce both higher flare rates and higher-energy events~\citep{walkowicz2011white}. 
The difference between the Solar and stellar-class FFDs is not only in the slope of the FFD but also in the normalisation.
For example, a flare of energy $10^{32}\,\mathrm{erg}$ is 3-6 orders of magnitude more frequent in other stellar types compared to the Sun as can be seen in~Fig.\,\ref{fig:flare_freq_energy_all_stellar_types}.
The low energy output from these stars however suffers from detection thresholds with some observations showing a flattening of the FFD at energies below $\leq 10^{30}\,\mathrm{erg}$~\citep{yang2019flare}.
In our stellar-class-dependent model, we adopt the FFD parameters and energy ranges from \citet{yang2019flare}, shown in Fig.\,\ref{fig:flare_freq_energy_all_stellar_types} and listed in Tab\,\ref{tab:stellar_FFED_params}. 
These parameters provide an observationally motivated estimate of the flare output from different stellar populations.

%
%

%


%
\subsection{Flare duration}\label{subsec:flare_duration}
In addition to the FFD, the duration of individual flares plays an important role in determining the quasi-persistent 511\,keV luminosity of stellar populations.
Observations of Solar flares indicate a positive correlation between flare duration and flare energy, with more energetic flares typically lasting longer.
For Solar flares, this relation has been quantified by~\citet{pettersen1989review}, who finds a power-law scaling of the form
%
%
\begin{equation}
    \log_{10}(\mathrm{t}_{0.5}) = 0.3\times\log_{10}(\mathrm{E_F}) - 7.5\mathrm{,}
\end{equation}
where $\mathrm{t}_{0.5}$ is the average decay time from maximum luminosity to half that level, and $\mathrm{E_F}$ is the total flare energy in erg. 
This implies that the flare duration scales approximately as $t \propto \mathrm{E_F^{0.3}}$.

More recent studies of stellar flares support a similar trend across a wide range of stellar types.
\citet{jackman2021NGTS} report a clear correlation between flare duration and bolometric energy for K- and M-type stars.
%
%
Similarly,~\citet{aegerter2020detection} find consistent scaling relations for Solar flares across different GOES classes.
While the exact normalization and slope may vary between datasets and stellar types, the general trend of increasing duration with energy appears consistent.
%
%
In our model, we adopt a power-law scaling between flare duration and energy,
\begin{equation}
    \Delta\mathrm{t}(\mathrm{E_F, M_*}) = \mathrm{T}\left(\dfrac{\mathrm{E_F}}{\mathrm{E_A}}\right)^{\tau},
\end{equation}
where $\mathrm{T}$ is a normalization constant, $\mathrm{E_A}$ is the pivot energy and $\tau \approx 0.3$ is the power-law index motivated by observations.

\section{Predicted Galactic luminosity from stellar flares}\label{sec:luminosity_predictions}
\subsection{Single-star quasi-persistent luminosity}
Using the framework introduced in Sect.\ref{sec:hierarchical_model} together with the FFDs and flare-durations described in Sect.\ref{sec:star_groups}, we estimate the quasi-persistent 511\,keV luminosity expected from a single star of a specific stellar type using Eq.\,\ref{eq:quasi_persistent_all_stars}, resp. Eq.\,\ref{eq:qp_lumi}.
This is shown in Fig.\,\ref{fig:quasi_luminosity}. 
The parameters for the FFDs of different stellar types and the respective luminosities obtained from them are listed in Tab\,\ref{tab:stellar_FFED_params}.
 The estimated luminosity differs based on whether the 511\,keV luminosity from the Sun is calculated using sub-correlations~(Sect.\ref{subsec: indirect_fit}) or directly using the SMM data~(Sect.\ref{subsec: direct_fit}), resulting in a broad range of values.
This luminosity is sensitive to the maximum flare energy, so that G-type stars appear to have the strongest quasi-persistent 511\,keV luminosity compared to other stellar types.
Another limitation arises from the rarity of sufficiently energetic flares.
Although the flare luminosity increases with flare energy, the FFDs decline steeply with energy, approximately as $E_F^{-2}$, except for A type stars.
As a result, the integrated luminosity remains dominated by relatively modest flare energies unless large maximum flare energies are assumed.
Even though M-type stars have a broader flare energy range and a similar maximum energy bound for flares compared to K-type stars, the quasi-persistent 511\,keV luminosity is smaller for M-stars compared to K-stars due to a steeper FFD.
\begin{table}[h!]
\centering
\caption{Flare frequency-energy relation power law indices and flare energy ranges as obtained from~\citet{yang2019flare}. Also listed are the 511\,keV luminosity values shown in Fig.\,\ref{fig:quasi_luminosity}.}
\begin{tabular}{c | c  c | c | c  c}
\hline
\textbf{Type} & $\mathbf{E_{\min}}$ & $\mathbf{E_{\max}}$ & $\boldsymbol{\alpha_\nu}$ & \multicolumn{2}{c}{$\mathbf{L^{*}_{511,QP}}$}\\
& \multicolumn{2}{c|}{$[\mathrm{erg}]$}
& 
& Indirect
& Direct \\
\hline
M & $10^{31}$ & $10^{35}$ &  -2.09 & $1.7\times 10^{25}$ & $1.4\times10^{27}$ \\ \hline
K & $10^{32}$ & $10^{35}$ & -1.78 & $8.2\times 10^{25}$ & $1.8\times10^{28}$ \\ \hline
G & $10^{32}$ & $10^{37}$ & -1.96 & $7.6\times 10^{26}$ & $2.2\times10^{30}$ \\ \hline
F & $10^{34}$ & $10^{36}$ & -2.11 & $6.5\times 10^{25}$ & $6.8\times10^{28}$ \\ \hline
A & $10^{34}$ & $10^{36}$ & -1.12 & $1.0\times 10^{25}$ & $2.3\times10^{28}$ \\ 
\hline
\end{tabular}
\label{tab:stellar_FFED_params}
\end{table}
\subsection{Total Galactic luminosity}
\begin{figure}[h!]
    \centering
    \includegraphics[width=\columnwidth]{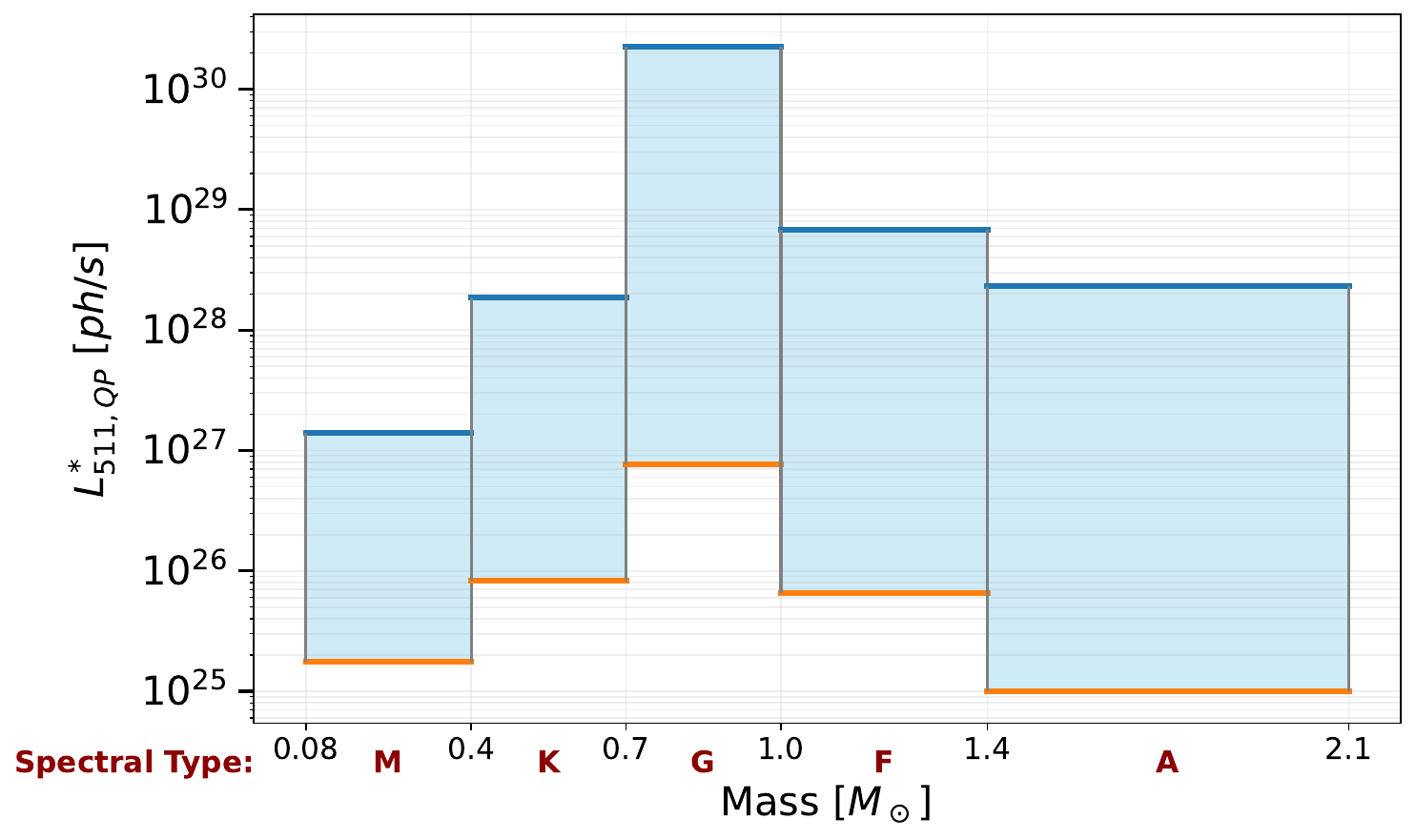}
    \caption{Quasi-persistent 511\,keV luminosity from a single star of a given stellar type. Orange line is the result of using parameters obtained in Sect.\,\ref{subsec: indirect_fit} from sub-correlations linking $\mathrm{E_F}, \mathrm{F_{SXR}}, \mathrm{F_{2223}}, \mathrm{F_{511}}$ and the blue line is from using the parameters from Sect.\,\ref{subsec: direct_fit} using the direct SMM based relation.}
    \label{fig:quasi_luminosity}
\end{figure}

To estimate the total Galactic positron annihilation luminosity expected from stellar flares, we integrate the quasi-persistent luminosity over the Galactic stellar population~(Eq.\ref{eq:luminosity_MW}).
As a first-order estimate, we assume a total of $\sim 10^{11}$ stars in the Milky Way and adopt a characteristic quasi-persistent luminosity of $\sim 10^{28}\,\mathrm{ph\,s^{-1}}$ per star (an optimistic approximation from Fig.\,\ref{fig:quasi_luminosity}).
This yields a total Galactic luminosity of
$\mathrm{L}_{511}^{\mathrm{tot}} \sim 10^{39}\,\mathrm{ph\,s^{-1}}$,
which is several orders of magnitude below the observed positron annihilation rate of $\sim 10^{43}\,\mathrm{e^+\,s^{-1}}$ inferred from INTEGRAL/SPI measurements.
Even allowing for uncertainties in the scaling relations and flare energies, this discrepancy cannot be bridged within reasonable parameter ranges. 
We refine this estimate by accounting for the stellar population more realistically.
%
%
Since the low-mass stars considered here have lifetimes comparable to the age of the Galaxy, we use the Kroupa initial mass function (IMF) of the Milky Way \citep{Kroupa2001_IMF} as an approximate PDMF to distribute the stars across stellar types and estimate their respective numbers by assuming the mass ranges given in Tab\,\ref{tab:imf_weighted_luminosity}.
%
%
Computing the total luminosity as a weighted sum of quasi-persistent 511\,keV luminosity across stellar types, we find a total Galactic luminosity of $\mathrm{L}_{511}^{\mathrm{tot}} \sim 10^{37}$--$10^{40}\,\mathrm{ph\,s^{-1}}$, where the lower limit corresponds to the indirect fitting of the sub-correlations and the upper limit corresponds to the fit to SMM data.
The parameters used to calculate the total 511\,keV luminosity in the Milky Way are listed in Tab.\,\ref{tab:imf_weighted_luminosity}.
\begin{table}
\centering
\caption{PDMF-weighted contribution of different stellar types to the Galactic quasi-persistent 511\,keV luminosity. The number fractions are computed using a Kroupa PDMF normalized over the mass range $0.08$--$2.1\,\mathrm{M_\odot}$ and assuming a total Galactic stellar population of $10^{11}$ stars. The luminosity range corresponds to the conservative and optimistic single-star quasi-persistent luminosities shown in Fig.\,\ref{fig:quasi_luminosity}.}
\label{tab:imf_weighted_luminosity}
\begin{tabular}{c c c c}
\hline
\textbf{Type} &
\textbf{Mass range} &
\textbf{\# in MW} &
\textbf{Contribution to } $\mathbf{L^{\rm MW}_{511}}$ \\
&
$[M_\odot]$ &
&
$[\mathrm{ph\,s^{-1}}]$ \\
\hline
M & $0.08$--$0.40$ &  $7.1\times10^{10}$ &  $1.2\times10^{36}$--$1.0\times10^{38}$ \\
K & $0.40$--$0.70$ &  $1.6\times10^{10}$ &  $1.3\times10^{36}$--$3.0\times10^{38}$ \\
G & $0.70$--$1.00$ &  $6.0\times10^{9}$ &  $4.6\times10^{36}$--$1.3\times10^{40}$ \\
F & $1.00$--$1.40$ &  $4.0\times10^{9}$ &  $2.3\times10^{35}$--$2.4\times10^{38}$ \\
A & $1.40$--$2.10$ &  $3.0\times10^{9}$ &  $2.7\times10^{34}$--$6.2\times10^{37}$ \\
\hline
Total & $0.08$--$2.10$ &  $1.0\times10^{11}$  & $7.4\times10^{36}$--$1.3\times10^{40}$ \\
\hline
\end{tabular}
\end{table}
%
%
Although energetics already disfavor the stellar flare scenario, morphology provides an independent and complementary test.
We therefore compare physically motivated stellar flare templates directly with INTEGRAL/SPI observations in the following.
%
\section{INTEGRAL/SPI Dataset and Analysis}\label{sec:INTEGRAL/SPI}

\subsection{Dataset}\label{subsec:SPI_dataset}
In this work, we analyze 20 years of INTEGRAL/SPI observations of the entire sky, selecting the energy range between 400--600\,keV in 6\,keV bins \citep[see][for a detailed explanation of the dataset]{yoneda2025imaging}. 
Our initial selection consisted of 143 833 pointed observations ("pointings") with an average observation time (dead-time corrected) of $\sim 2500$~s.
We filter out 206 individual pointings because no self-consistent background model could be created or the residuals in these observations were too large ($\pm10\sigma$).
The final dataset therefore has 143 627 pointing observations with a total gross exposure time of 367.4 Ms across the sky.
\subsection{Data Analysis}\label{subsec:SPI_data_analysis}
Raw SPI data is modeled by two general components in a linear combination: instrumental background and astrophysical sources.
The forward model reads
\begin{equation}
        d_{i,j,k} = \sum_{l} R_{l; ijk} \sum_{n=1}^{N_\mathrm{s}} \alpha_{nk} S_{nl} + \sum_{n=N_\mathrm{s}+1}^{N_\mathrm{s}+N_\mathrm{b}} \beta_{nk} B_{n;ijk}\mathrm{,}
        \label{eq:SPI_data_modeling}
    \end{equation}
where $d_{i,j,k}$ is the event counts, with $i, j,$ and $k$ being the indices of the pointing, detector, and energy bin that spans the data space, respectively.
$R_{l; ijk}$ is the instrument response for a given sky direction, $l$.
$\alpha_{nk}$ and $\beta_{nk}$ are the normalization factors (model parameters) for the $N_\mathrm{s}$ sky model components, $S_{n}$, and $N_\mathrm{b}$ background model components, $B_{n}$, respectively~\citep{Diehl2018_BGRDB}. 
The difference between the sky models and the background models is that the background does not depend on the observing direction of the instrument.
The background was created from the SPI background and response database \citep{Diehl2018_BGRDB, Siegert2019_SPIBG}. 
The background at a specific energy bin was modeled using two components: photons from continuum processes, and photons from $\gamma$-ray lines.
For an adequate fit, we find that a background re-normalization time scale of one parameter every INTEGRAL orbit ($\sim3$~d) for each of the two background components, is sufficient.
The total number of fitted background parameters is then 4556.
The sky parameters consist of 9 continuum point sources contributing to the 511\,keV bin, plus any additional models one would like to fit.
\subsection{Spatial stellar distribution model}\label{subsec:spatial_stellar_dist}
\begin{figure}
    \centering
    \includegraphics[width=0.95\linewidth]{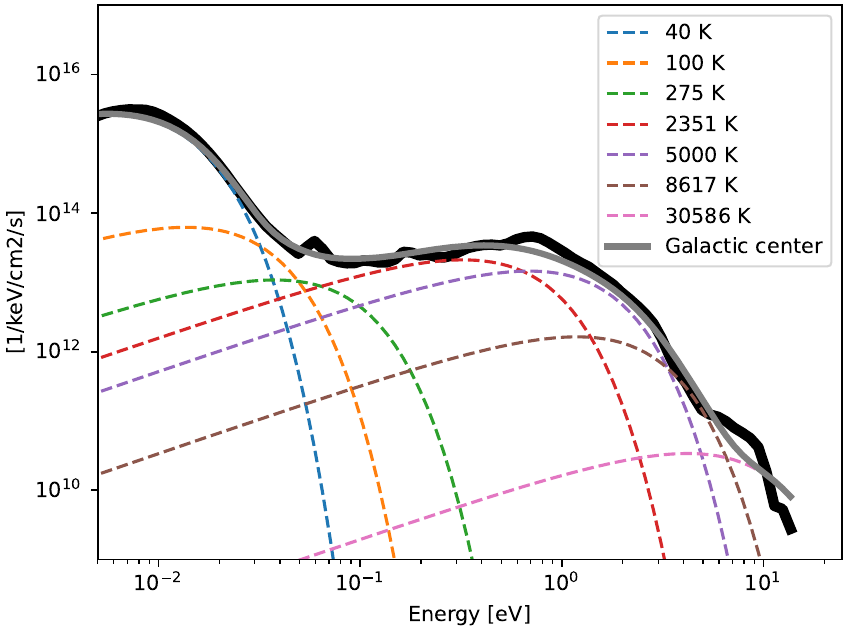}
    \caption{Blackbody decomposition of the total SED for the Galactic center voxel. The dashed lines mark the 7 blackbody temperatures used for different stellar types (O- to M- stars), and the grey line shows the combined emission compared to the modeled SEDs.}
    \label{fig:decomposition_blackbody}
\end{figure}

\begin{figure*}
    \centering
    \begin{subfigure}{0.33\linewidth}
        \centering
        \includegraphics[width=\linewidth]{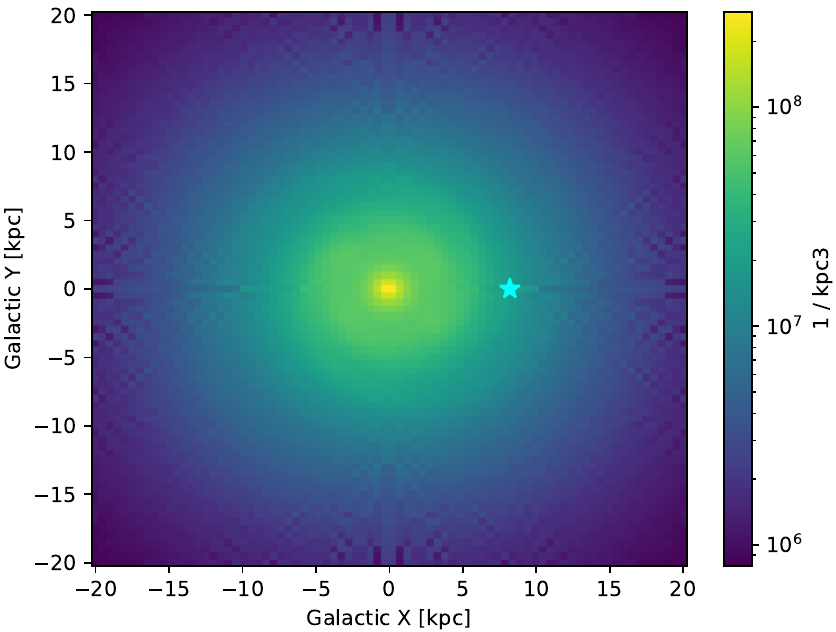}
        \caption{Distribution of M type stars}
    \end{subfigure}
    \begin{subfigure}{0.33\linewidth}
        \centering
        \includegraphics[width=\linewidth]{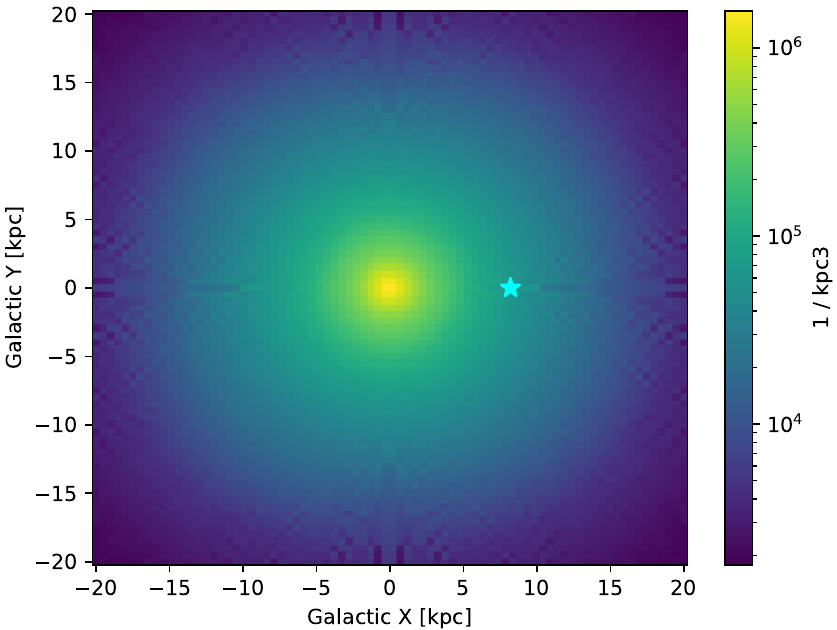}
        \caption{Distribution of G type stars}
    \end{subfigure}
    \begin{subfigure}{0.33\linewidth}
        \centering
        \includegraphics[width=\linewidth]{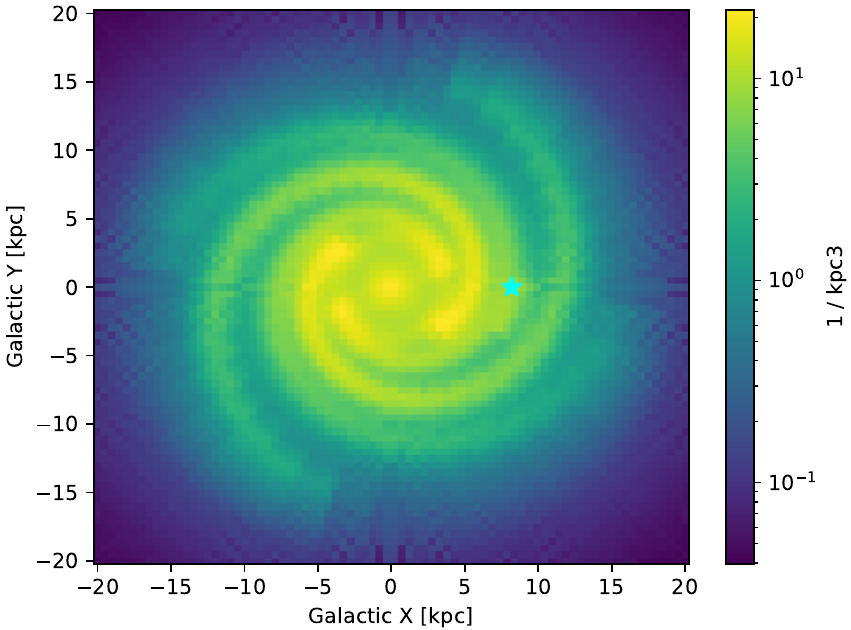}
        \caption{Distribution of B type stars}
    \end{subfigure}
    \caption{Number density of different stellar types in the Milky Way. The blue star is the position of the Sun.}
    \label{fig:six_plots}
\end{figure*}

\begin{figure*}
    \centering
    \begin{subfigure}{0.33\linewidth}
        \centering
        \includegraphics[width=\linewidth]{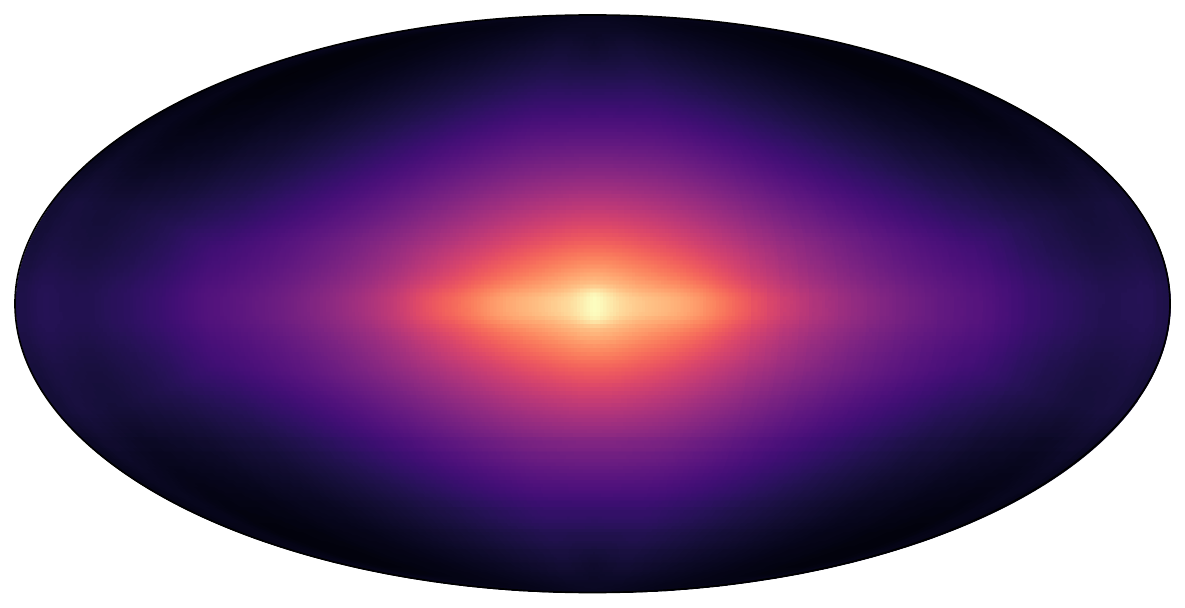}
        \caption{Flux map of M type stars}
    \end{subfigure}
    \begin{subfigure}{0.33\linewidth}
        \centering
        \includegraphics[width=\linewidth]{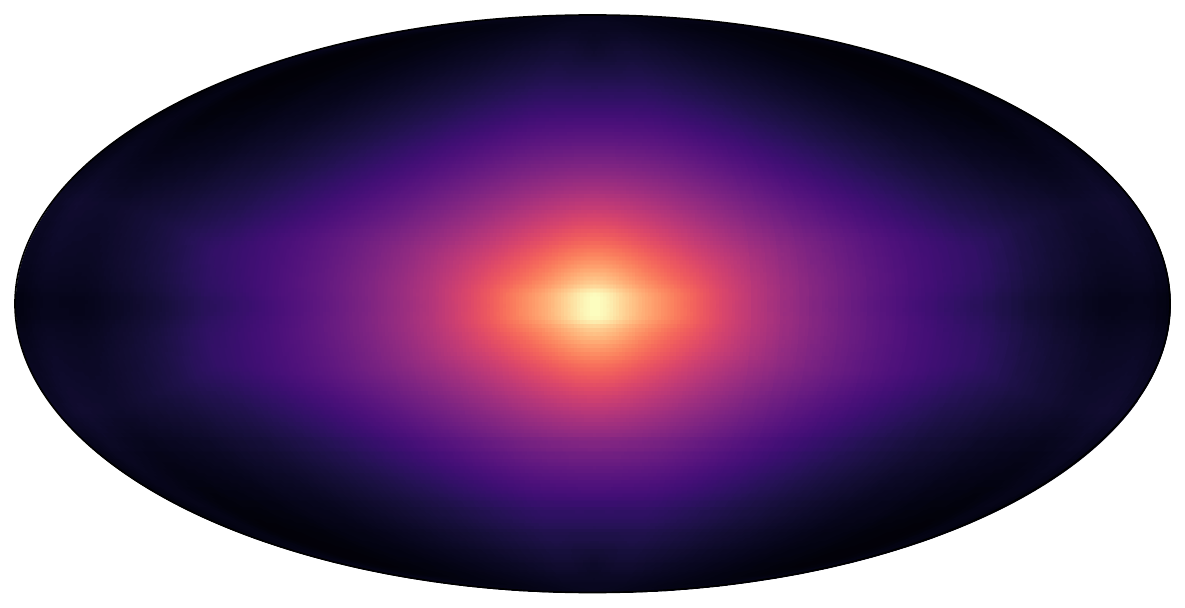}
        \caption{Flux map of G type stars}
    \end{subfigure}
%
%
    \begin{subfigure}{0.33\linewidth}
        \centering
        \includegraphics[width=\linewidth]{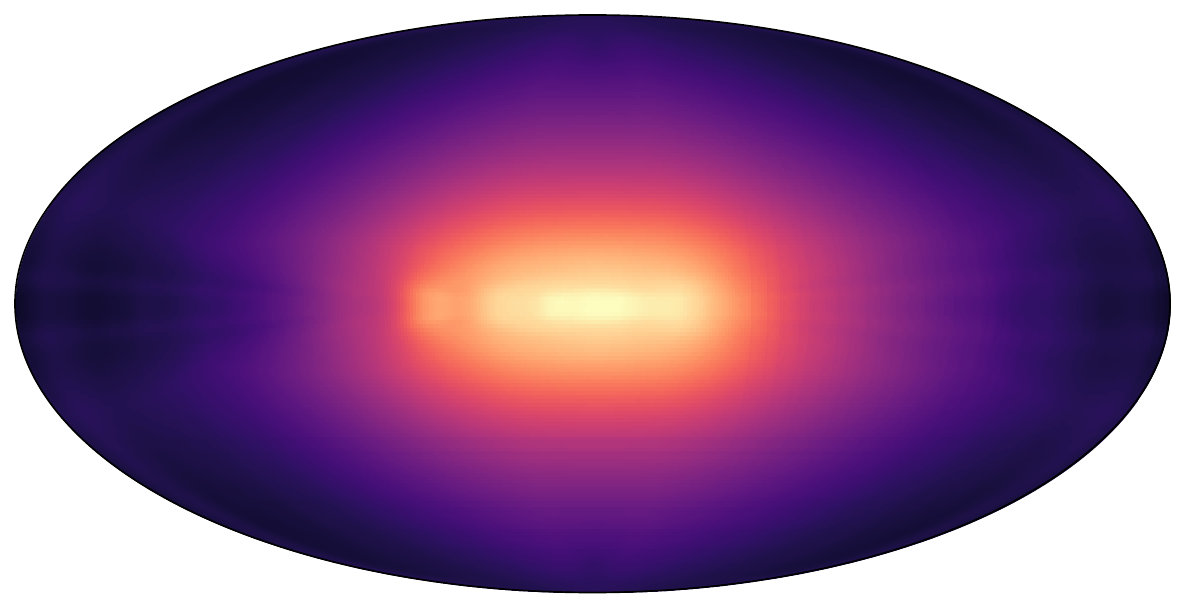}
        \caption{Flux map of B type stars}
    \end{subfigure}
%
    \caption{Line-of-sight integrated emissivity of the different stellar populations. Shown here is the distribution from 1\% (black) to 100\% (yellow/white) in logarithmic. All maps shown are in galactic coordinates and Mollweide projection.}
    \label{fig:LOS_six_plots}
\end{figure*}

\begin{figure}[!t]
    \centering
    \includegraphics[width=\columnwidth]{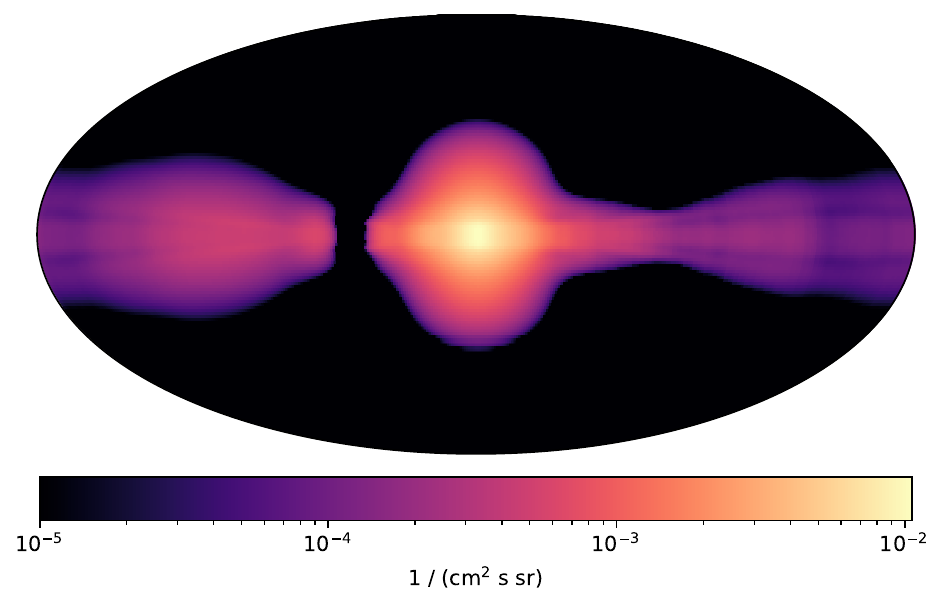}
    \caption{Fit to INTEGRAL/SPI data using M-, G-, B-, type stellar distribution templates.}
    \label{fig:MGB_fit}
\end{figure}

If stellar flares dominate Galactic positron production, the morphology of the 511\,keV emission should trace the underlying stellar populations responsible for the flare activity.
We therefore construct sky templates based on the Galactic distributions of different stellar types (O, B, A, F, G, K, M) and fit them directly to the SPI data.
%
%
%
%
We base our modeling on large-scale Galactic radiation field models~\citep{robitaille2017modular, porter2017high}, which provide the spectral energy distribution (SED) of the Milky Way as a function of position.
In each voxel, the total SED is decomposed into a linear combination of blackbody components representing different stellar types (O–M), each characterized by a fixed effective temperature.
The decomposition of the Galactic center voxel and the Solar neighborhood voxel is shown in Fig.\,\ref{fig:decomposition_blackbody}.
The resulting amplitudes of these blackbody components provide the relative contribution of each stellar type, and by extension an estimate of the number density of stars of each type per voxel.
This distribution is shown in Fig.\,\ref{fig:six_plots}.
Using the quasi-persistent 511\,keV luminosity per star shown in Fig.\,\ref{fig:quasi_luminosity}, we assign a corresponding emissivity to each stellar population.
Due to lack of observational data for flares from O- and B- type stars, we assume the same quasi-persistent 511\,keV luminosity as that of A- type stars.
The total 511\,keV emissivity is then obtained by summing over all stellar types.
Sky maps are generated by integrating the emissivity along the line of sight for each direction on the sky.
These maps are used as input templates in the SPI forward modeling (Sect.\,\ref{subsec:SPI_data_analysis}) and are shown in Fig.\,\ref{fig:LOS_six_plots}.
These distributions reflect the known structure of the Milky Way: young, massive stars (O, B and A types) are concentrated in the spiral arms and thin disk, while older, low-mass stars (G, K, and M types) are more broadly distributed and include a significant bulge and halo component.
For each stellar population, we assign a 511\,keV flux based on the stellar flare model (Sect.\,\ref{sec:luminosity_predictions}).
These templates are then fitted to the SPI data, either individually or in combination.
We find that individual stellar population templates, or one combined template with all stellar populations fails to reproduce the full morphology of the 511\,keV sky.
When using multiple templates with different stellar populations for the fit, significant degeneracies occur.
Different combinations of templates can produce similar structures making it difficult to uniquely disentangle the contributions of individual stellar types.
An interesting outcome is obtained when fitting the template maps of types= M, G, and B simultaneously in a single fit.
This model is able to reproduce both the bulge and disk components of the emission.
However, this results in some unphysical parameters in that it assigns negative fluxes to the B-type population.
The resulting skymap from this fit is shown in Fig.\,\ref{fig:MGB_fit}.

\subsection{Fitting globular clusters: eliminating the bulge}\label{subsec:glob_clusters_fit_SPI}
\begin{figure}[!t]
    \centering

    \begin{subfigure}{\columnwidth}
        \centering
        \includegraphics[width=\columnwidth]{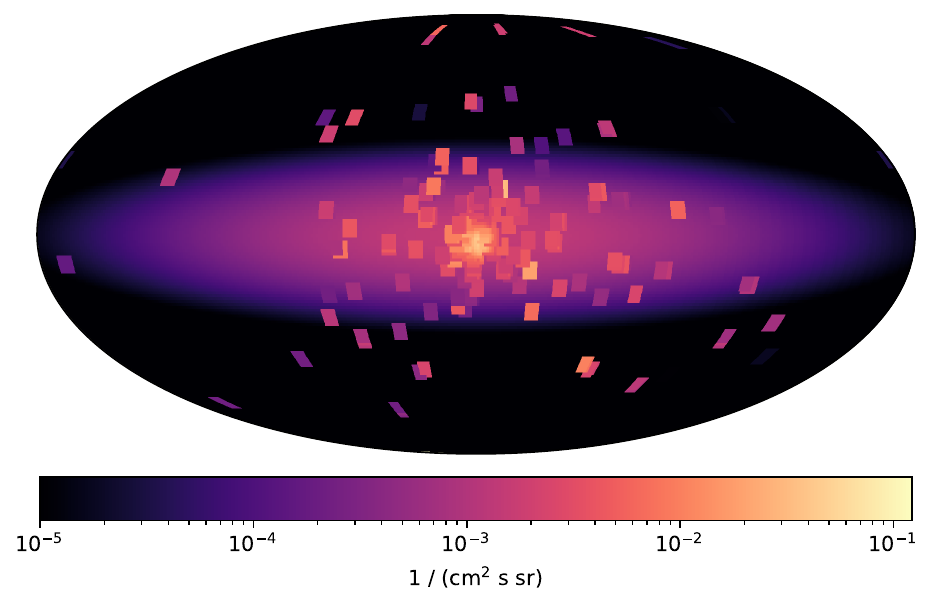}
        \caption{disk + NB + clusters ($\lambda=0$)}
        \label{fig:disk_nb_clusters_mass}
    \end{subfigure}

    \begin{subfigure}{\columnwidth}
        \centering
        \includegraphics[width=\columnwidth]{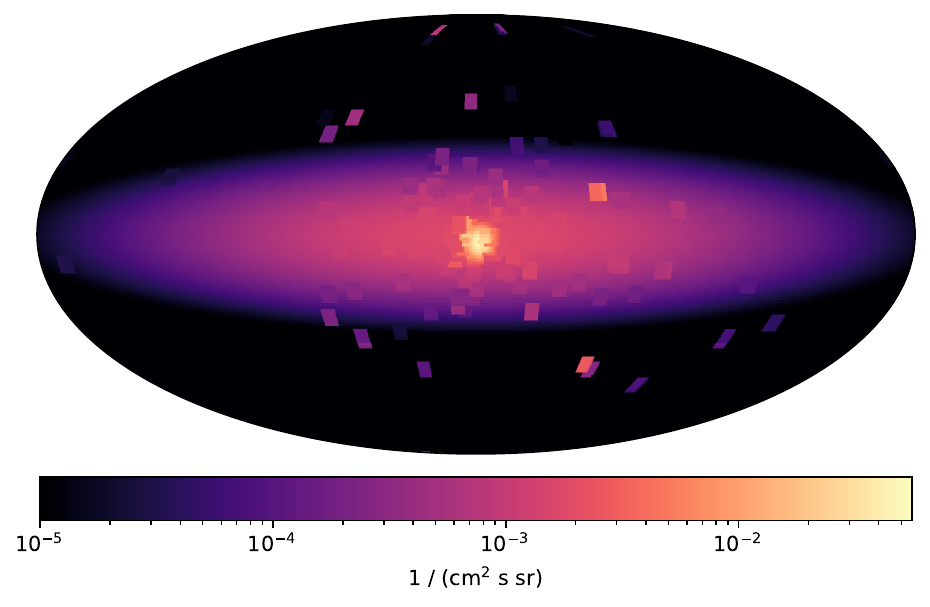}
        \caption{disk + NB + clusters (physical model)}
        \label{fig:disk_nb_clusters_physical}
    \end{subfigure}

    \caption{Comparison of disk + NB + clusters models.}
    \label{fig:disk_nb_clusters}
\end{figure}

Motivated by previous suggestions linking GCs and old stellar populations to Galactic positron production, and to better understand the broad bulge emission reported by~\citet{Siegert2016_511} and~\citet{yoneda2025imaging}, we test whether GCs can reproduce the observed bulge emission and potentially eliminate the need for an additional broad bulge component.
We construct a sky template based on the positions, distances, and masses of 154 known GCs of the Milky Way~\citep{baumgardt2018catalogue}.
Each cluster is treated as a point like source, and its expected flux is parametrized as:
\begin{equation}
    F_i \propto \dfrac{M_i^\lambda}{d_i^2},
\end{equation}
where $M_i$ and $d_i$ are the mass and distance of the $i$-th GC respectively, and $\lambda$ controls the dependence of the GC luminosity on the GC mass.
We explore a wide range of values for $\lambda$ from $[-10, 10]$, in steps of $0.5$, allowing for scenarios where massive clusters dominate $(\lambda > 0)$ or low mass clusters dominate $(\lambda < 0)$.
The case $\lambda = 0$ corresponds to all clusters having identical intrinsic luminosity, with only a distance dependence on the observed flux.
\\
Different GC templates in addition to a Galactic disk and narrow bulge (NB) component are incorporated into the SPI forward model (Eq.\,\ref{eq:SPI_data_modeling}).
The goal is to test whether the addition of GCs emission (agnostic of the stellar flare scenario) can replace the need for a broad bulge component.
\\
We find that none of the GC-based models provide an improved fit compared to the baseline model (disk, NB, and a broad bulge (BB))~\citep{Siegert2016_511}.
The likelihood values obtained for the GCs models are consistently worse ($\Delta\mathrm{log~likelihood} < -40$) than the baseline model, indicating that the spatial distribution of clusters cannot reproduce the extended bulge emission.
The best fit using the GCs model was obtained for the case of $\lambda = 0$, that is, when all GCs are assumed to have identical luminosities.
This may be considered unphysical as it ignores all dependence on the cluster mass or stellar content.
The best fit case for the disk, NB and GCs model $(\lambda = 0)$ is shown in Fig.\,\ref{fig:disk_nb_clusters_mass}. 
The likelihood values for all template fits are listed in Tab\,\ref{tab:log-likelihood}. \\
In a second step, we construct a physically motivated GC template by assigning cluster luminosities  based on our stellar flare model (Sect.\,\ref{sec:luminosity_predictions}).
Each individual cluster is assigned a total 511\,keV luminosity relative to one cluster.
This results in a more realistic prediction for the GCs contribution to the 511\,keV sky.
Comparing the results of this physically motivated GC template (together with the disk and NB components), no significant improvement over the disk + NB model alone was found.
The result of this fit is shown in Fig.\,\ref{fig:disk_nb_clusters_physical}.
This indicates that GCs when modeled with detailed stellar flare physics, do not significantly contribute to the observed 511\,keV emission.
This does not exclude GCs to be viable 511\,keV test objects as other sources, such as X-ray binaries or pulsars~\citep{Bartels2018_binaries511}, may be more frequent in different GCs.
%

%

\section{Discussion}\label{sec:discussion}
Our results disfavor stellar flares as the dominant source of the Galactic 511\,keV emission. 
First, the energetics are insufficient even under optimistic assumptions for the flare energy-511\,keV luminosity relation, the flare duration, and the FFD. 
The predicted Galactic luminosity remains several orders of magnitude below the \(\sim 10^{43} - 10^{44}\,\mathrm{e^+\,s^{-1}}\) annihilation rate inferred from INTEGRAL/SPI observations. 
Second, the spatial morphology does not provide evidence for a significant stellar-flare contribution. 
In particular, the physically motivated GC template, does not significantly improve the fit compared to the disk plus narrow bulge model.

Several sources of uncertainties may affect our analysis.
The calibration of the $\mathrm{E_F}$-$\mathrm{L_{511}}$ relation relies on Solar observations, which may not fully capture the behavior of more active stars.
Additionally, the intermediate correlations used to establish this relation might introduce some uncertainties that could not fully be accounted for in the fitting process.
The FFDs for different stellar types are also subject to observational biases especially at low energies where detection limits become significant.
Despite these uncertainties, the discrepancy between predicted and observed luminosities span several orders of magnitude, making it unlikely that systematic effects alone could bridge this gap between this scenario and observations.

%
Our model assumes that the physical processes governing positron production in Solar flares extend to other stellar types.
While this is a reasonable first order approximation, differences in magnetic field strengths, stellar rotations and other physical conditions could modify flare behavior and particle acceleration mechanisms.
Additionally, our treatment of stellar populations is simplified relying on average properties and scaling relations.

Our calculation is calibrated to the 511\,keV luminosity from positrons that annihilate locally in the flare.
If a fraction of flare-produced positrons escapes the stellar atmosphere, it could in principle contribute to the total 511\,keV luminosity predicted from stellar flares.
However, plausible escape fractions at the level $f_{\mathrm{esc}} = 0.2$ would only modify our predictions by 25\% assuming all escaped positrons annihilate in the interstellar medium.
In contrast, the assumption from~\citet{Bisnovatyi-Kogan2017_511}, where only $\sim2\times10^{-6}$ of the produced positrons annihilate locally, implies that $99.9998\%$ escape and would boost our predictions to $\mathrm{L_{511}} \sim 10^{46}~\mathrm{ph\,s^{-1}}$, further suggesting that the earlier assumed escape fraction is highly optimistic.
%
%

The stellar flare scenario could in principle be enhanced if one considers extreme assumptions 
such as increasing the maximum flare energy for all stellar types to higher values.
To account for 10\% of the Galactic 511\,keV luminosity, one would need to reach $10^{39}\,\mathrm{erg}$ as the maximum flare energy of all stars with a continuous FFD. 
If the maximum flare energy of all stellar types reaches up to $\gtrsim 10^{37 - 39}~\mathrm{erg}$, the stellar flare model would account for $3.8 \times 10^{40} - 2.0 \times 10^{42}~\mathrm{ph\,s^{-1}}$ in 511\,keV.
This is currently not supported by observations, and neither are enhanced flare rates across all stellar populations.
Other parameters such as flare-duration, and the ratio of the 511\,keV energy to the bolometric flare energy could affect the estimates.
The quasi-persistent luminosity is linear in both the flare duration and the flare rate, therefore, if flares were systematically ten times longer than assumed in our current model, the predicted luminosity would increase by an order of magnitude.
Similarly, a larger 511\,keV-to-bolometric energy ratio in other stars could enhance the prediction, but this would require stellar flares to be much more efficient positron producers than Solar flares, for which there is currently no direct observational calibration.
%
However, such adjustments still do not resolve the mismatch in spatial morphology.

Another possible candidate to enhance the flaring star scenario are flaring brown dwarfs. 
The number of brown dwarfs in the Milky Way is estimated to be between $10^{10}$ and $10^{11}$~\citep{carnero2019brown}.
While not all brown dwarfs are expected to flare, certainly a fraction of L and T brown dwarfs do~\citep{xin2024huge}.
The FFDs of these objects are unknown, but assuming a similar trend as for M dwarfs might enhance the quasi-persistent 511\,keV flux by a factor of 2.
Likewise, our model is gauged to a total number of Milky Way stars of $10^{11}$. 
Newer censuses of the Galactic stellar population may range more up to $2 \times 10^{11}$~\citep{licquia2015improved} or even $4 \times 10^{11}$~\citep{nasa_blueshift_2015_milkyway_stars}.
This would enhance the luminosity by another factor of 2–4, which may, in total, lead to a luminosity of up to $10^{41}\,\mathrm{ph\,s^{-1}}$ in 511\,keV in the most optimistic case.

Such a value may be tested by observations of the 2.223\,MeV line. 
Since the correlation of 511\,keV flux to the 2.223\,MeV flux in Solar flares is basically linear ($\mathrm{F_{511}} = (0.65 \pm 0.07) \times \mathrm{F_{2223}}^{1.22\pm0.07}$), one could expect a quasi-diffuse 2.223\,MeV line with a similar distribution as the 511\,keV line in the Milky Way.
\citet{mcconnell1997comptel} searched for the 2.223\,MeV line using CGRO/COMPTEL but, apart from a tentative 4$\sigma$ point-like signal away from the Galactic bulge and disk~\citep{mcconnell1997possible}, there was no flux detected up to a level of $(1-2) \times 10^{-5}\,\mathrm{ph\,cm^{-2}\,s^{-1}}$ for point sources, which can be converted to an upper limit on a diffuse-like flux from the Galaxy of about $10^{-4}\,\mathrm{ph\,cm^{-2}\,s^{-1}}$.
With more than 22\,yr of INTEGRAL/SPI observations, and thanks to the development of a flexible and reliable background model for the high-energy range of SPI above 2\,MeV~\citep{weinberger2021supernova, Siegert2022_MWdiffuse}, a diffuse emission search for the 2.223\,MeV line is within reach.
Given the upper bound on the 511\,keV line luminosity of $10^{41}\,\mathrm{ph\,s^{-1}}$, which converts into a Galactic-wide flux of $10^{-5}\,\mathrm{ph\,cm^{-2}\,s^{-1}}$ from stellar flares only, a similar 2.223\,MeV line flux would be expected. 
Using the $^{26}\mathrm{Al}$ all-sky analysis from \citet{Pleintinger2023_26Al} as an estimate for the uncertainty of diffuse emission around 2\,MeV, we expect a sensitivity of $2 \times 10^{-5}\,\mathrm{ph\,cm^{-2}\,s^{-1}}$ for the full sky, and $3 \times 10^{-6}\,\mathrm{ph\,cm^{-2}\,s^{-1}}$ for the Galactic ridge. 
A future search for the 2.223\,MeV will provide additional constraints for this stellar flare scenario. 

Based on our fits to individual GCs, we can compare several interesting sources which have literature predictions to our upper limits in Tab.\,\ref{tab:fit_fluxes_165}: Terzan\,5, Liller\,1, NGC\,6440, and NGC\,6441 \citep{Bartels2018_binaries511}.
The Galactic X-ray binary scenario from \citet{Bartels2018_binaries511} also suggests GCs as test case, where several objects should show 511\,keV flux on the order of $10^{-5}\,\mathrm{ph\,cm^{-2}\,s^{-1}}$.
Terzan\,5 should therefore be the strongest GC and should show a flux of $2.9 \times 10^{-5}\,\mathrm{ph\,cm^{-2}\,s^{-1}}$, whereas our $2\sigma$ upper limit is right at $<2.7 \times 10^{-5}\,\mathrm{ph\,cm^{-2}\,s^{-1}}$.
Likewise, Liller\,1 is expected at a level of $2.2 \times 10^{-5}\,\mathrm{ph\,cm^{-2}\,s^{-1}}$, while our limit is at $<3.8 \times 10^{-5}\,\mathrm{ph\,cm^{-2}\,s^{-1}}$.
NGC\,6440 and 6441 would be found around $1.0 \times 10^{-5}\,\mathrm{ph\,cm^{-2}\,s^{-1}}$, for which our limits are $< 4.2$ and $<3.3 \times 10^{-5}\,\mathrm{ph\,cm^{-2}\,s^{-1}}$, respectively.
While, formally, Terzan\,5 excludes the binary scenario on the $2\sigma$ level, there is still some room for uncertainty in the modeling as well as in the observations.

In the case of $\omega$\,Centauri (NGC\,5139), the most massive GC, our upper limit of $<3.5 \times 10^{-5}\,\mathrm{ph\,cm^{-2}\,s^{-1}}$ converts to a luminosity limit of $< 10^{41}\,\mathrm{ph\,s^{-1}}$ which is hardly constraining in itself due to the large distance.
However, it is again two orders of magnitude stronger than radio recombination line searches, setting limits on the luminosity of $< 10^{43}\,\mathrm{ph\,s^{-1}}$ \citep[$3\sigma$;][]{Staveley-Smith2022_Psalpha}.
If the combined limit of the closest and most constraining GC is used, which is here the case of NGC\,6838 with $< 6 \times 10^{40}\,\mathrm{ph\,s^{-1}}$ and no source confusion, several other scenarios, such as dark-matter-related expectations, could be investigated.
A discussion about all GCs is beyond the scope of this paper.
However clearly, INTEGRAL/SPI reaches the realm of extremely deep annihilation observations to search for protruding 511\,keV point sources which may be the smoking gun for this long-standing conundrum.

Nevertheless, stellar flares may still contribute at a subdominant level. 
Positrons from stellar flares could add to contributions from radioactive isotopes, compact binaries, millisecond pulsars, type Ia supernovae, or other old-population sources. 
Such a contribution is physically plausible due to the large number of flaring stars.
Additionally, the concentration of a large number of GCs in the bulge region might contribute to the boxy bulge emission reducing the need to transport positrons from the disk to the bulge.

\section{Summary and outlook}\label{sec:conclusion}
In this work, we have investigated the contribution of stellar flares to the Galactic 511\,keV emission using a hierarchical modeling framework calibrated on Solar observations.
We find that stellar flares produce insufficient positrons by several orders of magnitude to explain the observed Galactic signal contrary to the earlier study by~\citet{Bisnovatyi-Kogan2017_511} who suggested that all positrons in the bulge can be explained by stellar flares.
Reproducing the required positron injection rate would require extremely large flare energies.
Spatial models based on stellar populations fail to reproduce the morphology of the 511\,keV emission.
Stellar flare based models of globular clusters do not significantly improve agreement with observations.
These results strongly disfavor stellar flares as the dominant source of Galactic positrons.

The Galactic 511\,keV emission is therefore more likely to originate from a combination of source populations.
Radioactive isotopes particularly $^{26}\mathrm{Al}$ produced by massive stars provide an established contribution to the Galactic disk but the bright bulge may be accounted for by using old stellar populations such as compact binaries, millisecond pulsars or type Ia supernovae.
Additional contribution could potentially be related with past activity in the Galactic center or more exotic processes such as dark matter.
Although our results disfavor stellar flares as the dominant positron source in globular clusters, the clusters themselves cannot yet be excluded as potential sources because they contain high number of old stars and compact objects and are concentrated around the Galactic bulge.
A better understanding of positron propagation through the interstellar medium may further bridge the gap between the injection sites and the observed annihilation morphology.

Next generation $\gamma$-ray instruments such as NASA's Compton Spectrometer and Imager (COSI) Small Explorer mission, which will be launched in 2027~\citep{Tomsick2019_COSI,Tomsick2024}, will play a crucial role in advancing our understanding of Galactic positrons.
COSI with its improved spectral resolution, wide field of view and imaging capabilities in the MeV regime, is expected to provide significantly enhanced measurements of the 511\,keV line and associated continuum emission.
Future work with detailed treatment of positron transport in the interstellar medium will be essential for identifying the origin of Galactic positrons and resolving this  longstanding astrophysical puzzle.
Finally, a diffuse 2.223\,MeV $\gamma$-ray line search with INTEGRAL/SPI and in the future with COSI will shed more light on the stellar flare scenario due to the strong correlation of the 511\,keV and 2.223\,MeV line in Solar flares.

\begin{acknowledgements}
Saurabh Mittal acknowledges support by the  {Bundesministerium für Wirtschaft und Energie via the Deutsches Zentrum für Luft- und Raumfahrt (DLR)} under Contract No. {50\,OO\,2219}. 
Laura Eisenberger acknowledges support by the Bundesministerium für Wirtschaft und Energie via the Deutsches Zentrum für Luft- und Raumfahrt (DLR) under Contract No. {50\,OR\,2413} and is grateful for the support of the {Stu\-di\-en\-stif\-tung des Deu\-tschen Vol\-kes}. 
Dimitris Tsatsis and Tristan Bouchet acknowledge support from the {DFG/LIS} project {SI\,2502/6-1, project number 551127478}. 
Rudi Reinhardt acknowledges support by the \emph{Bun\-des\-mi\-nis\-te\-ri\-um für Forschung, Technologie und Raumfahrt} as part of the \emph{FORKA (Forschung für den Rückbau kerntechnischer Anlagen)} funding program under the reference number {15S9455\,A-E}.
T.O. was supported by JSPS Overseas Research Fellowships.
FHP is supported by a Forrest Research Foundation Fellowship. 
HY is supported by JSPS KAKENHI Grant Number 23K13136 and 26H00833.
\end{acknowledgements}

\bibliographystyle{aa} 
\bibliography{saurabh} 

\appendix
\section{}\label{app:A}
\subsection{bulge region: random point source distribution}\label{subsec:bulge_random}
\begin{figure}[!h]
    \centering
    \includegraphics[width=\columnwidth]{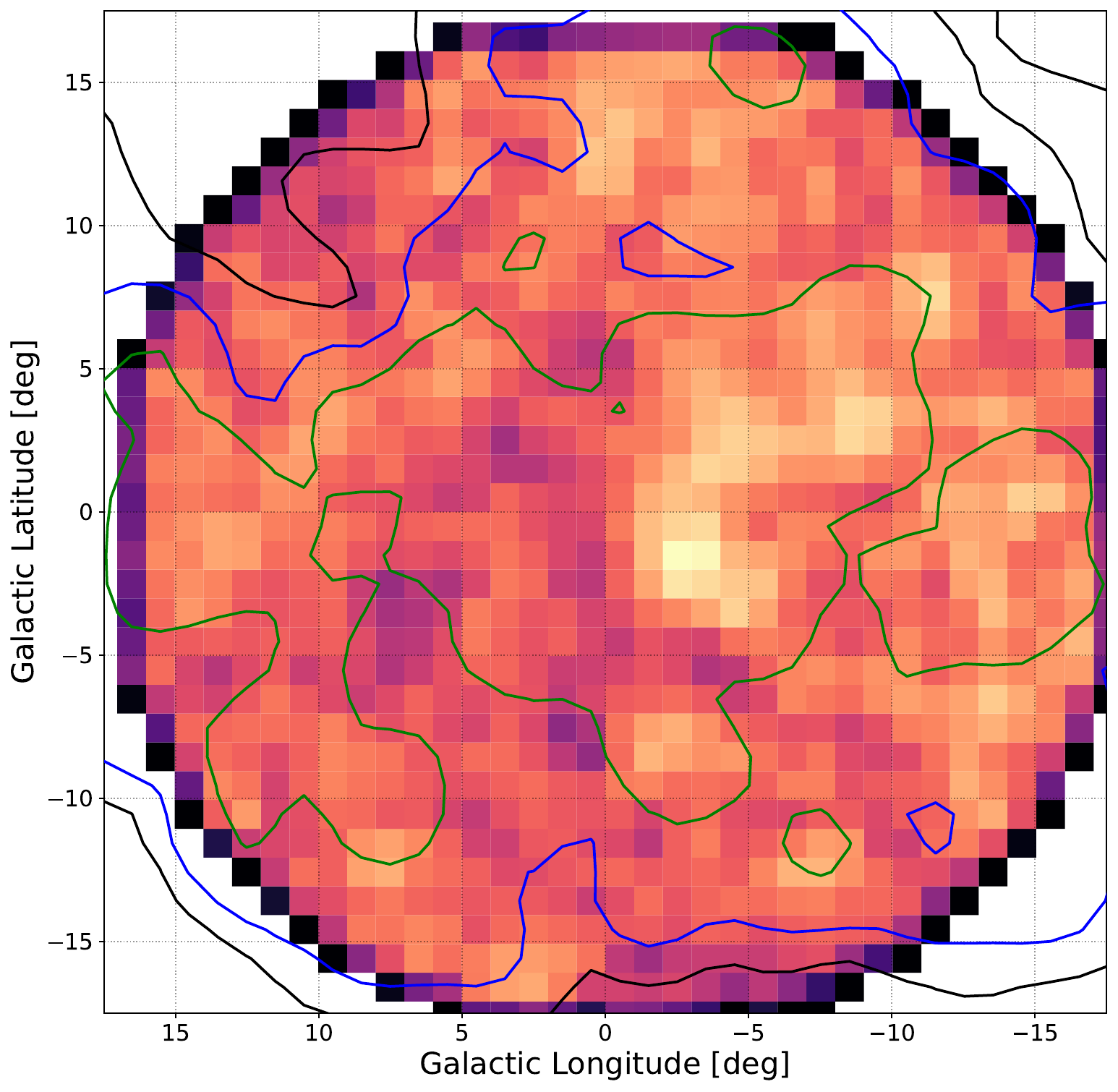}
    \caption{Average of the 107 best-fit random point-source distributions within $35^\circ$ of the Galactic center. Overlaid contours show the 90th, 95th, and 99th percentile levels derived from the image reproduced from~\citet{yoneda2025imaging}, shown in green, blue, and black, respectively.}
    \label{fig:bulge_random}
\end{figure}

Given that GCs fail to reproduce the bulge morphology, we next investigate whether the observed emission could arise from a more general population of unresolved point sources associated with old stars~\citep{Knoedlseder2005_511, Weidenspointner2008_511, Bartels2018_GeVexcess_stars}.
We construct different template models in which the bulge is represented by a set of randomly distributed 150 point sources within $35^\circ$ of the Galactic Center.
%
%
We test 1200 of these realizations along with a disk and NB component.
We find that the random point source model, in some cases (107 out of 1200), can approximate the bulge emission reasonably well, resulting in a better likelihood than the baseline model.
The quality of the fit varies significantly between different realizations indicating a strong preference for a certain spatial configuration of sources.
The average of the 107 best-fit realizations shows an asymmetry and substructure as shown in Fig.\,\ref{fig:bulge_random}.
This suggests that the true 511\,keV emission in the bulge is not smooth and cannot be fully explained by a simple parametric model such as a Gaussian bulge.
%
%
\subsection{Clusters as individual point sources}\label{subsec:clusters_individual}
\begin{figure}[!]
    \centering
    \includegraphics[width=\columnwidth]{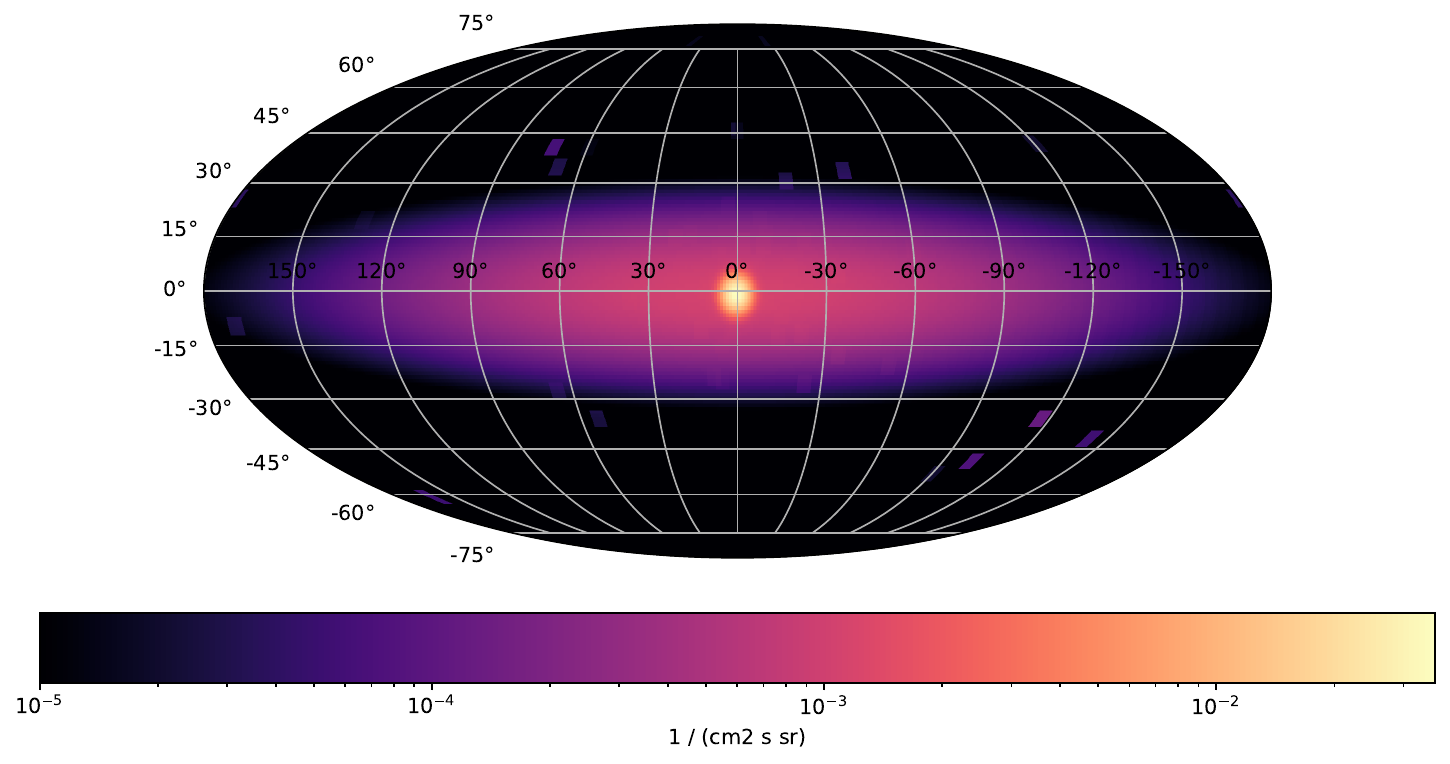}
    \caption{Skymap obtained from fitting a disk, NB, 9 bright sources from the SPI catalog, and 154 GCs as individual point sources to the INTEGRAL/SPI data.}
    \label{fig:clusters_individual}
\end{figure}

\begin{figure}[!]
    \centering
    \begin{subfigure}{0.49\linewidth}
        \centering
        \includegraphics[width=\linewidth]{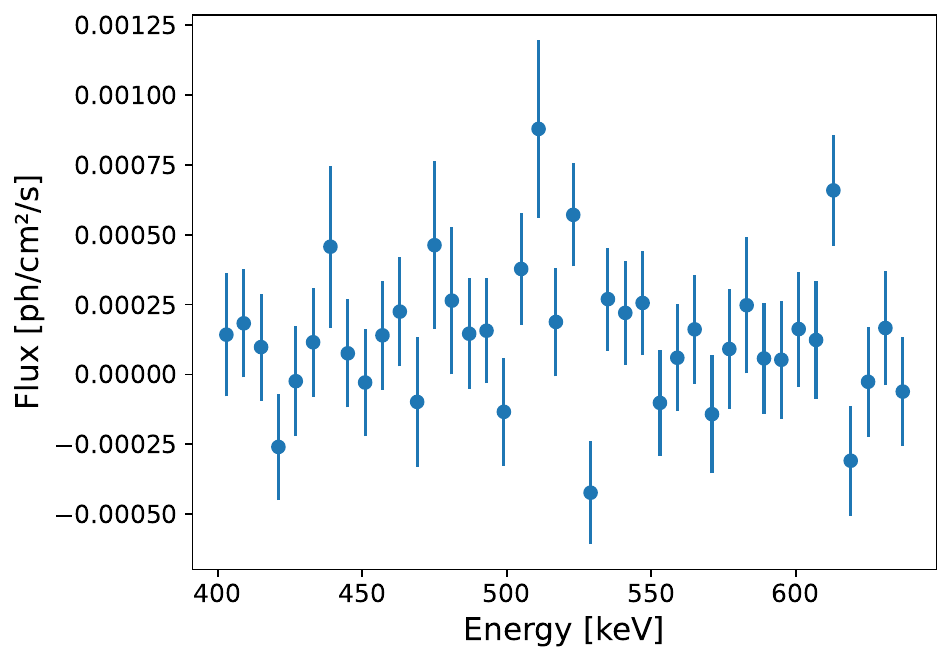}
        \caption{154 globular clusters}
    \end{subfigure}
    \hfill
    \begin{subfigure}{0.49\linewidth}
        \centering
        \includegraphics[width=\linewidth]{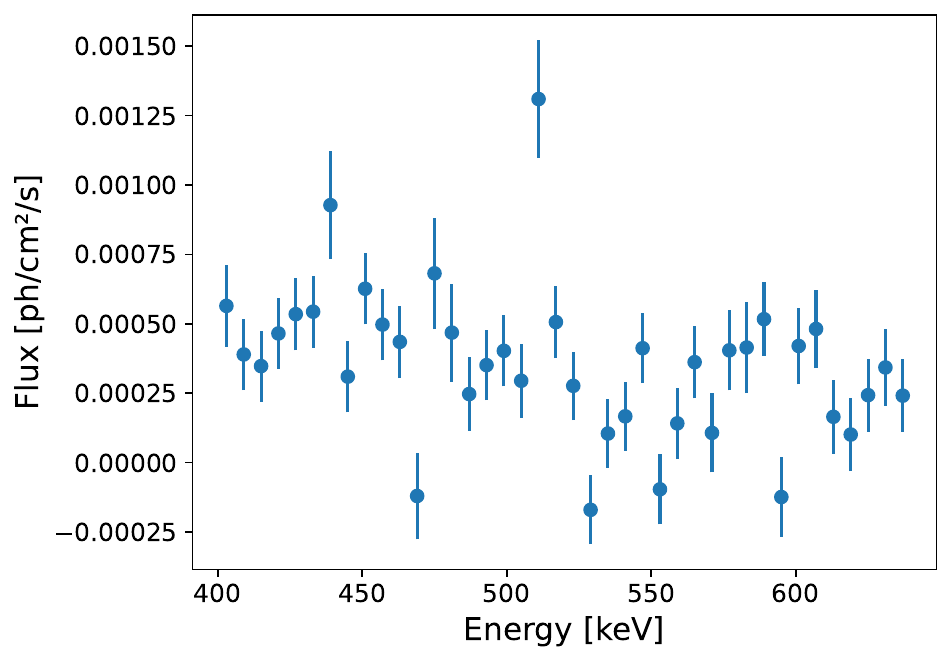}
        \caption{Disk}
    \end{subfigure}
    \hfill
    
    \vspace{0.5cm}

    \begin{subfigure}{0.49\linewidth}
        \centering
        \includegraphics[width=\linewidth]{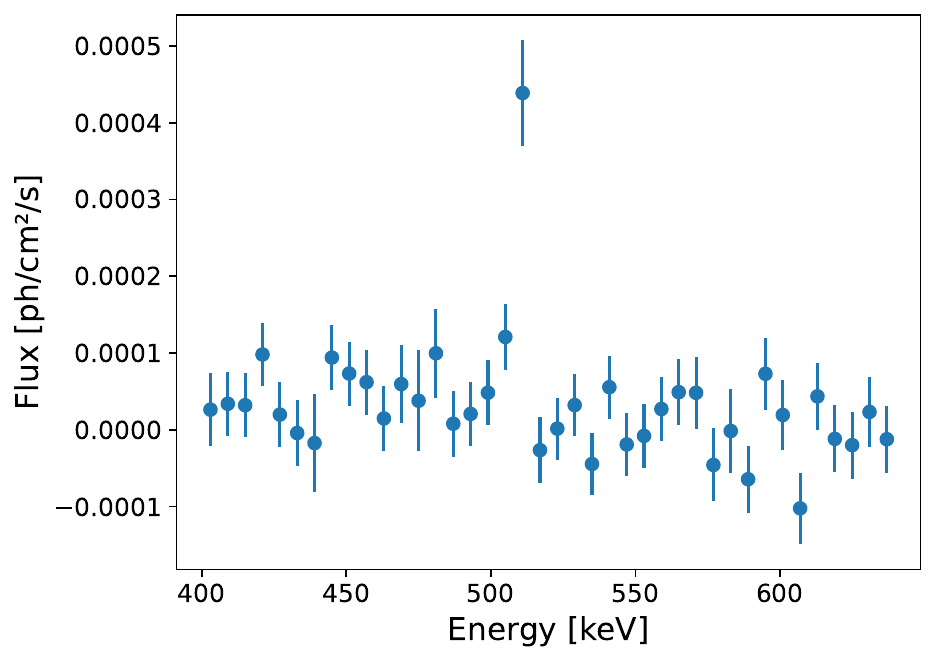}
        \caption{Narrow bulge}
    \end{subfigure}
    \hfill
    \begin{subfigure}{0.49\linewidth}
        \centering
        \includegraphics[width=\linewidth]{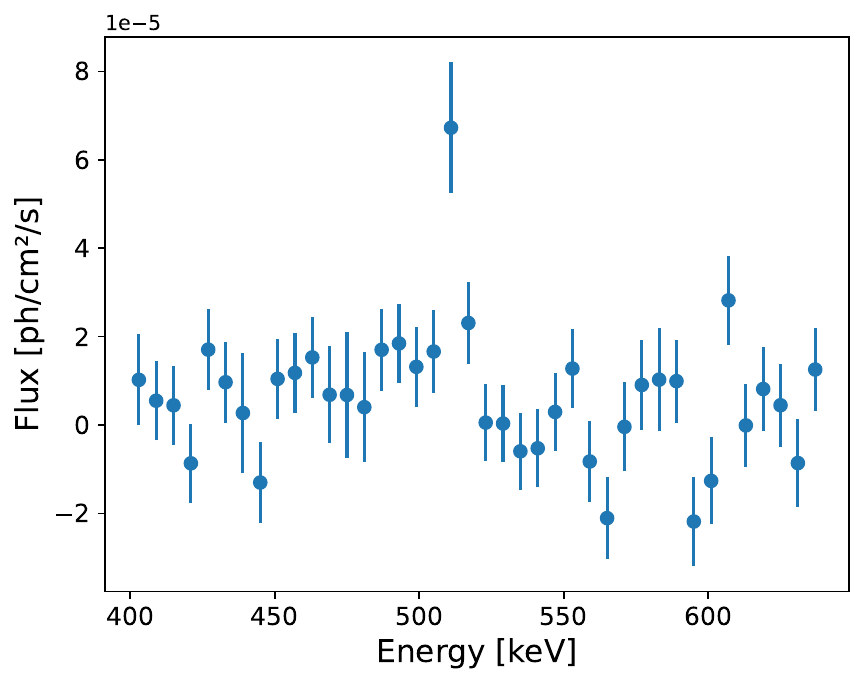}
        \caption{Galactic center source}
    \end{subfigure}
    \hfill

    \caption{Spectrum of different components obtained from fitting to the INTEGRAL/SPI data using a disk, NB, 9 bright sources from the SPI catalog, and 154 GCs as individual point sources.}
    \label{fig:fit_spectra_indiv_fit}
\end{figure}

Instead of using the GCs as a template component, we fit them as individual point sources.
In total, we fit a disk component, NB, 154 GCs and 9 bright point sources from the SPI source catalog.
The result of this fit is shown in Fig.\,\ref{fig:clusters_individual}.
Some clusters do receive a significant flux from this fit.
This does not automatically mean that these clusters can be resolved as individual point sources.
This could merely be a result of some emission seen in those regions~\citep{yoneda2025imaging}, and attributed to the clusters coinciding within the region.
The spectrum for some of the components from this fit namely, disk, NB, Galactic Center Source, and the combined spectrum from all GCs is shown in Fig.\,\ref{fig:fit_spectra_indiv_fit}.

\begin{table}[!]
\centering
\caption{Different template fits and their respective log-likelihood in the 511\,keV bin. All log-likelihood values are with respect to the Baseline fit (disk + NB + BB), where negative values are worse fits than the baseline. Each fit uses 4556 background components and 9 bright sources from the SPI catalog.}
\scriptsize
\renewcommand{\arraystretch}{0.90}
\begin{tabular}{c | c | c}
\hline
\textbf{Fit components} & \textbf{\# of fit params} & \textbf{log-likelihood}\\
\hline
M stars & 4566 & -344.80\\ \hline
K or G stars & 4566 & -192.72\\ \hline
F or A stars & 4566 & -280.35\\ \hline
B or O stars & 4566 & -609.02\\ \hline
M, G, B template & 4568 & -54.29\\ \hline
Disk + NB & 4567 & -53.98 \\ \hline
Disk + NB + clusters ($\alpha = 0$) & 4568 & -42.45\\ \hline
Disk + NB + clusters (physical model) & 4568 & -53.87 \\ \hline
Disk + NB + clusters (point source) & 4721 & 76.95\\ \hline
Map~\citet{yoneda2025imaging} & 4566 & 77.71\\ \hline

\end{tabular}
\label{tab:log-likelihood}
\end{table}
\begin{table*}[!t]
\centering
\caption{Best-fit fluxes for the disk, NB, nine bright SPI sources, and 154 GCs (sorted by distance). Entries with $z=F/\sigma_F>2$ are shown as $F\pm\sigma_F$; otherwise, the $2\sigma$ upper limit is reported. The dagger marks the nine bright SPI continuum sources; their quoted values correspond to the 508--514 keV bin and is not the 511\,keV line flux. Asterisks mark sources within $1.5^\circ$ of another listed source (source confusion: values and limits are potentially over- or under-estimated).}
\label{tab:fit_fluxes_165}
\scriptsize
\setlength{\tabcolsep}{2.2pt}
\renewcommand{\arraystretch}{0.8}
\begin{tabular}{l c c c c @{\hspace{0.25cm}}|@{\hspace{0.25cm}} l c c c c}
\hline
Component & $l$ & $b$ & $d$ & Flux & Component & $l$ & $b$ & $d$ & Flux \\
 & [deg] & [deg] & [kpc] & [$10^{-5}\,\mathrm{ph\,cm^{-2}\,s^{-1}}$]& & [deg] & [deg] & [kpc] & [$10^{-5}\,\mathrm{ph\,cm^{-2}\,s^{-1}}$]\\
\hline
Disk & -- & -- & $5.0$--$15.0$ & $131.0 \pm 21.2$ & NGC 6637$^{*}$ & 1.72 & -10.27 & 8.8 & $<5.2$ \\
Narrow bulge & -- & -- & $\sim 8.0$ & $43.9 \pm 6.9$ & NGC 6093 & 352.67 & 19.46 & 8.9 & $<5.4$ \\
Crab$^{\dagger}$ & 184.55 & -5.79 & 2.0 & $13.5 \pm 1.4$ & NGC 6144$^{*}$ & 351.93 & 15.70 & 8.9 & $<6.2$ \\
Cygnus X$-$1$^{\dagger}$ & 71.33 & 3.06 & 2.2 & $2.6 \pm 1.2$ & NGC 5927$^{*}$ & 326.60 & 4.86 & 9.1 & $5.7 \pm 2.1$ \\
PSR B1509-58$^{\dagger}$ & 320.31 & -1.17 & 5.0 & $<4.7$ & NGC 362 & 301.53 & -46.25 & 9.2 & $<4.8$ \\
Galactic center source$^{\dagger}$ & 359.94 & -0.05 & 8.2 & $6.7 \pm 1.5$ & NGC 6402 & 21.32 & 14.80 & 9.3 & $<7.6$ \\
GRS 1915+105$^{\dagger}$ & 45.36 & -0.22 & 8.6 & $<3.9$ & NGC 6681 & 2.85 & -12.51 & 9.3 & $<4.8$ \\
Swift J1753.5$-$0127$^{\dagger}$ & 24.90 & 12.18 & 8.8 & $<4.3$ & NGC 6287 & 0.13 & 11.02 & 9.4 & $<2.9$ \\
Centaurus A$^{\dagger}$ & 309.51 & 19.42 & $3 \cdot 10^3$ & $<3.6$ & NGC 5272 & 42.22 & 78.71 & 9.6 & $<6.9$ \\
MR 2251-178$^{\dagger}$ & 46.17 & -61.35 & $4 \cdot 10^5$ & $<26.1$ & NGC 6779 & 62.66 & 8.34 & 9.7 & $<2.9$ \\
3C 273$^{\dagger}$ & 290.00 & 64.34 & $7 \cdot 10^5$ & $<2.7$ & NGC 6380$^{*}$ & 350.18 & -3.42 & 9.8 & $<7.5$ \\
NGC 6121$^{*}$ & 350.97 & 15.97 & 1.9 & $<7.4$ & NGC 6139 & 342.37 & 6.94 & 9.8 & $<4.7$ \\
NGC 6397 & 338.17 & -11.96 & 2.4 & $4.5 \pm 1.8$ & FSR 1735 & 339.19 & -1.85 & 9.8 & $<2.7$ \\
NGC 6544$^{*}$ & 5.84 & -2.20 & 2.6 & $<3.3$ & NGC 6293 & 357.62 & 7.83 & 9.8 & $<2.7$ \\
NGC 6656$^{*}$ & 9.89 & -7.55 & 3.2 & $<3.4$ & NGC 288 & 151.29 & -89.38 & 10.0 & $<27.2$ \\
Terzan 12 & 8.36 & -2.10 & 3.4 & $<4.6$ & NGC 6652$^{*}$ & 1.53 & -11.38 & 10.0 & $<3.6$ \\
2MASS-GC01$^{*}$ & 10.47 & 0.10 & 3.6 & $3.6 \pm 1.6$ & NGC 4590 & 299.63 & 36.05 & 10.1 & $<8.6$ \\
NGC 6366 & 18.41 & 16.04 & 3.7 & $<5.2$ & NGC 2808 & 282.19 & -11.25 & 10.2 & $<3.6$ \\
NGC 6838 & 56.75 & -4.56 & 4.0 & $<2.9$ & NGC 7078 & 65.01 & -27.31 & 10.2 & $<9.7$ \\
NGC 6752 & 336.49 & -25.63 & 4.2 & $<9.4$ & NGC 6638 & 7.90 & -7.15 & 10.3 & $<5.2$ \\
NGC 104 & 305.89 & -44.89 & 4.4 & $<4.7$ & NGC 7089 & 53.37 & -35.77 & 10.5 & $<9.9$ \\
NGC 3201 & 277.23 & 8.64 & 4.6 & $<3.1$ & NGC 5986 & 337.02 & 13.27 & 10.6 & $<5.2$ \\
NGC 6218 & 15.72 & 26.31 & 4.7 & $<6.5$ & NGC 6569$^{*}$ & 0.48 & -6.68 & 10.6 & $<6.6$ \\
NGC 6254 & 15.14 & 23.08 & 5.0 & $<6.0$ & NGC 6517 & 19.23 & 6.76 & 10.6 & $<4.6$ \\
NGC 6540$^{*}$ & 3.29 & -3.31 & 5.2 & $<4.1$ & NGC 5946$^{*}$ & 327.58 & 4.19 & 10.6 & $<4.1$ \\
NGC 5139 & 309.10 & 14.97 & 5.2 & $<3.5$ & NGC 6388 & 345.56 & -6.74 & 10.7 & $<5.3$ \\
NGC 6809 & 8.79 & -23.27 & 5.3 & $<6.3$ & NGC 2298 & 245.63 & -16.01 & 10.8 & $<8.5$ \\
IC 1276 & 21.83 & 5.67 & 5.4 & $<3.7$ & Palomar 1 & 130.06 & 19.03 & 11.0 & $<7.7$ \\
NGC 6626 & 7.80 & -5.58 & 5.4 & $<2.7$ & NGC 6496 & 348.03 & -10.01 & 11.3 & $<4.1$ \\
Terzan 5 & 3.84 & 1.69 & 5.5 & $<2.7$ & NGC 1851 & 244.51 & -35.04 & 11.3 & $12.1 \pm 5.5$ \\
NGC 4372 & 300.99 & -9.88 & 5.8 & $<4.1$ & NGC 5286 & 311.61 & 10.57 & 11.4 & $<3.2$ \\
NGC 6304$^{*}$ & 355.83 & 5.38 & 5.8 & $<4.3$ & NGC 6453 & 355.72 & -3.87 & 11.6 & $<3.9$ \\
Palomar 6 & 2.09 & 1.78 & 5.8 & $<3.1$ & NGC 6316$^{*}$ & 357.18 & 5.76 & 11.6 & $<4.3$ \\
NGC 6522$^{*}$ & 1.02 & -3.93 & 5.8 & $<10.4$ & NGC 6441 & 353.53 & -5.01 & 11.8 & $<3.3$ \\
NGC 6352 & 341.42 & -7.17 & 5.9 & $<3.4$ & NGC 5897 & 342.95 & 30.29 & 12.6 & $<7.1$ \\
Palomar 10 & 52.44 & 2.72 & 5.9 & $<2.7$ & Palomar 8 & 14.10 & -6.80 & 12.8 & $<5.0$ \\
NGC 6171 & 3.37 & 23.01 & 6.0 & $<5.2$ & NGC 6584 & 342.14 & -16.41 & 13.2 & $<4.5$ \\
NGC 4833 & 303.60 & -8.02 & 6.2 & $<4.5$ & NGC 1904 & 227.23 & -29.35 & 13.3 & $<9.2$ \\
Djorgovski 2$^{*}$ & 2.76 & -2.51 & 6.3 & $<4.6$ & NGC 6235 & 358.92 & 13.52 & 13.5 & $3.8 \pm 1.5$ \\
Ton 2$^{*}$ & 350.80 & -3.42 & 6.4 & $<4.6$ & Djorgovski 1 & 356.67 & -2.48 & 13.7 & $<4.8$ \\
NGC 6624 & 2.79 & -7.91 & 6.4 & $<2.9$ & Palomar 11 & 31.81 & -15.58 & 14.3 & $<3.9$ \\
NGC 6256 & 347.79 & 3.31 & 6.4 & $<3.5$ & NGC 6356$^{*}$ & 6.72 & 10.22 & 15.1 & $<4.6$ \\
NGC 6266 & 353.57 & 7.32 & 6.4 & $<4.0$ & NGC 6284 & 358.35 & 9.94 & 15.1 & $<5.2$ \\
NGC 6535 & 27.18 & 10.44 & 6.5 & $<5.0$ & NGC 6934 & 52.10 & -18.89 & 15.4 & $<5.2$ \\
BH 261 & 3.36 & -5.27 & 6.5 & $<2.6$ & NGC 1261 & 270.54 & -52.12 & 15.5 & $<9.2$ \\
ESO 452-SC11 & 351.91 & 12.10 & 6.5 & $3.0 \pm 1.5$ & NGC 6101 & 317.75 & -15.82 & 16.1 & $<5.0$ \\
Terzan 4$^{*}$ & 356.02 & 1.31 & 6.7 & $<3.6$ & NGC 5466 & 42.15 & 73.59 & 16.9 & $<6.0$ \\
Terzan 1$^{*}$ & 357.56 & 0.99 & 6.7 & $<5.2$ & NGC 6981 & 35.16 & -32.68 & 17.0 & $<6.1$ \\
Terzan 6 & 358.57 & -2.16 & 6.7 & $<4.2$ & NGC 5053$^{*}$ & 335.70 & 78.95 & 17.2 & $<8.0$ \\
NGC 6553$^{*}$ & 5.25 & -3.03 & 6.8 & $<3.3$ & NGC 5024$^{*}$ & 332.96 & 79.76 & 17.9 & $<6.3$ \\
NGC 6205 & 59.01 & 40.91 & 6.8 & $<6.6$ & NGC 4147 & 252.85 & 77.19 & 18.2 & $<3.2$ \\
HP 1$^{*}$ & 357.43 & 2.12 & 6.8 & $<6.4$ & IC 4499 & 307.35 & -20.47 & 18.2 & $<7.5$ \\
NGC 6712 & 25.35 & -4.32 & 7.0 & $<4.1$ & Palomar 12 & 30.51 & -47.68 & 19.0 & $<10.6$ \\
Terzan 9$^{*}$ & 3.60 & -1.99 & 7.1 & $<6.3$ & Rup 106 & 300.89 & 11.67 & 21.2 & $<5.7$ \\
2MASS-GC02$^{*}$ & 9.78 & -0.62 & 7.1 & $<3.3$ & NGC 6864 & 20.30 & -25.75 & 21.6 & $<5.7$ \\
NGC 6717 & 12.88 & -10.90 & 7.1 & $4.6 \pm 1.6$ & Terzan 7 & 3.39 & -20.07 & 22.8 & $<3.8$ \\
NGC 6558$^{*}$ & 0.20 & -6.02 & 7.2 & $<4.1$ & ESO 280-SC06 & 346.90 & -12.57 & 22.9 & $<6.9$ \\
NGC 6362 & 325.55 & -17.57 & 7.4 & $<9.0$ & Palomar 5 & 0.85 & 45.86 & 23.2 & $<11.6$ \\
NGC 6528$^{*}$ & 1.14 & -4.17 & 7.5 & $<13.5$ & NGC 6715 & 5.61 & -14.09 & 24.1 & $<4.9$ \\
Terzan 2$^{*}$ & 356.32 & 2.30 & 7.5 & $<4.8$ & Palomar 13 & 87.10 & -42.70 & 24.8 & $<8.6$ \\
FSR 1716 & 329.78 & -1.59 & 7.5 & $<4.4$ & IC 1257 & 16.53 & 15.14 & 25.0 & $<6.4$ \\
NGC 5904 & 3.86 & 46.80 & 7.6 & $<11.3$ & NGC 7492 & 53.39 & -63.48 & 26.6 & $<14.8$ \\
NGC 6401 & 3.45 & 3.98 & 7.7 & $<3.6$ & Terzan 8 & 5.76 & -24.56 & 26.7 & $<5.2$ \\
UKS 1 & 5.13 & 0.76 & 7.8 & $<3.2$ & Palomar 2 & 170.53 & -9.07 & 27.2 & $<5.0$ \\
NGC 6749 & 36.20 & -2.20 & 7.8 & $<5.0$ & NGC 5634 & 342.21 & 49.26 & 27.2 & $<7.4$ \\
NGC 6325 & 0.97 & 8.00 & 7.8 & $<2.6$ & Arp 2 & 8.55 & -20.79 & 28.6 & $<4.6$ \\
NGC 6539 & 20.79 & 6.78 & 7.8 & $<4.3$ & NGC 6229 & 73.64 & 40.31 & 30.6 & $6.3 \pm 3.1$ \\
NGC 6541 & 349.29 & -11.19 & 8.0 & $<3.4$ & Whiting 1 & 161.62 & -60.64 & 31.3 & $<11.3$ \\
NGC 6760 & 36.11 & -3.92 & 8.0 & $<2.9$ & NGC 5824 & 332.56 & 22.07 & 31.8 & $<4.2$ \\
Terzan 10$^{*}$ & 4.49 & -1.99 & 8.0 & $<5.0$ & AM 4 & 320.28 & 33.51 & 32.2 & $<8.5$ \\
Lynga 7 & 328.77 & -2.80 & 8.0 & $<3.9$ & NGC 6426 & 28.09 & 16.23 & 35.3 & $<4.8$ \\
NGC 7099 & 27.18 & -46.84 & 8.0 & $<10.7$ & NGC 5694 & 331.06 & 30.36 & 37.3 & $<5.2$ \\
NGC 6642$^{*}$ & 9.81 & -6.44 & 8.1 & $<4.7$ & Pyxis & 261.32 & 7.00 & 39.4 & $<3.6$ \\
E 3 & 292.27 & -19.02 & 8.1 & $<4.6$ & NGC 7006 & 63.77 & -19.41 & 42.8 & $<6.7$ \\
Liller 1 & 354.84 & -0.16 & 8.1 & $<3.8$ & Palomar 15 & 18.85 & 24.34 & 45.1 & $<6.7$ \\
Terzan 3 & 345.08 & 9.19 & 8.1 & $<3.2$ & Palomar 14 & 28.75 & 42.19 & 71.0 & $<10.1$ \\
NGC 6440 & 7.73 & 3.80 & 8.2 & $<4.2$ & NGC 2419 & 180.37 & 25.24 & 83.2 & $<30.9$ \\
NGC 6273 & 356.87 & 9.38 & 8.3 & $<3.0$ & Eridanus & 218.11 & -41.33 & 90.1 & $<18.3$ \\
NGC 6723 & 0.07 & -17.30 & 8.3 & $<3.4$ & Palomar 3 & 240.14 & 41.86 & 92.5 & $<15.3$ \\
NGC 6333$^{*}$ & 5.54 & 10.71 & 8.4 & $<3.9$ & Palomar 4 & 202.31 & 71.80 & 103.0 & $<7.0$ \\
NGC 6342$^{*}$ & 4.90 & 9.73 & 8.4 & $<4.8$ & AM 1 & 258.36 & -48.47 & 123.3 & $<16.6$ \\
NGC 6341 & 68.34 & 34.86 & 8.4 & $<8.7$ & Crater & 274.81 & 47.85 & 145.0 & $<8.6$ \\
NGC 6355 & 359.58 & 5.43 & 8.7 & $<2.3$ &  &  &  &  &  \\
\hline
\end{tabular}
\end{table*}

\end{document}